\pdfoutput=1
\documentclass[12pt,a4paper]{article}

\usepackage{ifthen} % for conditional statements
\newboolean{pdflatex}
\setboolean{pdflatex}{true} % False for eps figures 

\newboolean{articletitles}
\setboolean{articletitles}{true} % False removes titles in references

\newboolean{uprightparticles}
\setboolean{uprightparticles}{false} %True for upright particle symbols

\newboolean{inbibliography}
\setboolean{inbibliography}{false} %True once you enter the bibliography

\def\paperauthors{LHCb collaboration} % Leave as is for PAPER and CONF
\def\paperasciititle{Measurement of Ds production asymmetry in pp collisions at sqrt(s)=7 and 8 TeV} % Set ASCII title here
\def\papertitle{Measurement of \Dspm production asymmetry in $pp$ collisions at $\sqrt{s}=7$ and $8\tev$} % Latex formatted title
\def\paperkeywords{{High Energy Physics}, {LHCb},{flavour physics},{forward physics},{charm physics},{heavy quark production}} % Comma separated list
\def\papercopyright{\the\year\ CERN for the benefit of the LHCb collaboration} % new since 9/Apr/2018
\def\paperlicence{CC-BY-4.0 licence}
\def\paperlicenceurl{https://creativecommons.org/licenses/by/4.0/}

\usepackage[top=1in, bottom=1.25in, left=1in, right=1in]{geometry}

\usepackage{booktabs}
\usepackage[british]{babel} % to ensure correct hyphenation
\usepackage{microtype}
\usepackage{lineno}  % for line numbering during review
\usepackage{xspace} % To avoid problems with missing or double spaces after
\usepackage{caption} %these three command get the figure and table captions automatically small

\usepackage{graphicx}  % to include figures (can also use other packages)
\usepackage{color}
\usepackage{colortbl}
\graphicspath{{./figs/}} % Make Latex search fig subdir for figures

\usepackage{amsmath} % Adds a large collection of math symbols
\usepackage{amssymb}
\usepackage{amsfonts}
\usepackage{upgreek} % Adds in support for greek letters in roman typeset

\newcommand*\patchAmsMathEnvironmentForLineno[1]{%
\expandafter\let\csname old#1\expandafter\endcsname\csname #1\endcsname
\expandafter\let\csname oldend#1\expandafter\endcsname\csname
end#1\endcsname
 \renewenvironment{#1}%
   {\linenomath\csname old#1\endcsname}%
   {\csname oldend#1\endcsname\endlinenomath}%
}
\newcommand*\patchBothAmsMathEnvironmentsForLineno[1]{%
  \patchAmsMathEnvironmentForLineno{#1}%
  \patchAmsMathEnvironmentForLineno{#1*}%
}
\AtBeginDocument{%
\patchBothAmsMathEnvironmentsForLineno{equation}%
\patchBothAmsMathEnvironmentsForLineno{align}%
\patchBothAmsMathEnvironmentsForLineno{flalign}%
\patchBothAmsMathEnvironmentsForLineno{alignat}%
\patchBothAmsMathEnvironmentsForLineno{gather}%
\patchBothAmsMathEnvironmentsForLineno{multline}%
\patchBothAmsMathEnvironmentsForLineno{eqnarray}%
}

\usepackage{hyperxmp}

\usepackage[pdftex,
            pdfauthor={\paperauthors},
            pdftitle={\paperasciititle},
            pdfkeywords={\paperkeywords},
            pdfcopyright={Copyright (C) \papercopyright},
            pdflicenseurl={\paperlicenceurl}]{hyperref}

\usepackage[all]{hypcap} % Internal hyperlinks to floats.

\usepackage{xspace} 
\usepackage{upgreek}

\def\lhcb {\mbox{LHCb}\xspace}

\def\lhc    {\mbox{LHC}\xspace}

\def\MagUp {\mbox{\em Mag\kern -0.05em Up}\xspace}
\def\MagDown {\mbox{\em MagDown}\xspace}

\ifthenelse{\boolean{uprightparticles}}%
{
 
 \def\Pgamma      {\ensuremath{\upgamma}\xspace}

 \def\Pmu         {\ensuremath{\upmu}\xspace}

 \def\Ppi         {\ensuremath{\uppi}\xspace}

 \def\Pphi        {\ensuremath{\upphi}\xspace}

 \def\Ppsi        {\ensuremath{\uppsi}\xspace}

 \def\PDelta      {\ensuremath{\Delta}\xspace}                 
 \def\PXi      {\ensuremath{\Xi}\xspace}                 
 \def\PLambda      {\ensuremath{\Lambda}\xspace}                 
 \def\PSigma      {\ensuremath{\Sigma}\xspace}                 
 \def\POmega      {\ensuremath{\Omega}\xspace}                 
 \def\PUpsilon      {\ensuremath{\Upsilon}\xspace}                 
 
 \def\PB      {\ensuremath{\mathrm{B}}\xspace}                 
                  
 \def\PD      {\ensuremath{\mathrm{D}}\xspace}

 \def\PJ      {\ensuremath{\mathrm{J}}\xspace}                 
 \def\PK      {\ensuremath{\mathrm{K}}\xspace}

 \def\Pb      {\ensuremath{\mathrm{b}}\xspace}                 
 \def\Pc      {\ensuremath{\mathrm{c}}\xspace}

 \def\Pi      {\ensuremath{\mathrm{i}}\xspace}

 \def\Pp      {\ensuremath{\mathrm{p}}\xspace}

 \def\Ps      {\ensuremath{\mathrm{s}}\xspace}

}
{
 
 \def\Pgamma      {\ensuremath{\gamma}\xspace}

 \def\Pmu         {\ensuremath{\mu}\xspace}

 \def\Ppi         {\ensuremath{\pi}\xspace}

 \def\Pphi        {\ensuremath{\phi}\xspace}

 \def\Ppsi        {\ensuremath{\psi}\xspace}                 
                  
 \mathchardef\PDelta="7101
 \mathchardef\PXi="7104
 \mathchardef\PLambda="7103
 \mathchardef\PSigma="7106
 \mathchardef\POmega="710A
 \mathchardef\PUpsilon="7107
                  
 \def\PB      {\ensuremath{B}\xspace}                 
                  
 \def\PD      {\ensuremath{D}\xspace}

 \def\PJ      {\ensuremath{J}\xspace}                 
 \def\PK      {\ensuremath{K}\xspace}

 \def\Pb      {\ensuremath{b}\xspace}                 
 \def\Pc      {\ensuremath{c}\xspace}

 \def\Pi      {\ensuremath{i}\xspace}

 \def\Pp      {\ensuremath{p}\xspace}

 \def\Ps      {\ensuremath{s}\xspace}

}

\makeatletter
\ifcase \@ptsize \relax% 10pt
  \newcommand{\miniscule}{\@setfontsize\miniscule{4}{5}}% \tiny: 5/6
\or% 11pt
  \newcommand{\miniscule}{\@setfontsize\miniscule{5}{6}}% \tiny: 6/7
\or% 12pt
  \newcommand{\miniscule}{\@setfontsize\miniscule{5}{6}}% \tiny: 6/7
\fi
\makeatother

\DeclareRobustCommand{\optbar}[1]{\shortstack{{\miniscule (\rule[.5ex]{1.25em}{.18mm})}
  \\ [-.7ex] $#1$}}

\def\mup        {{\ensuremath{\Pmu^+}}\xspace}
\def\mun        {{\ensuremath{\Pmu^-}}\xspace} % muon negative (\mum is taken)

\def\g      {{\ensuremath{\Pgamma}}\xspace}

\def\squark    {{\ensuremath{\Ps}}\xspace}

\def\cquark    {{\ensuremath{\Pc}}\xspace}
\def\cquarkbar {{\ensuremath{\overline \cquark}}\xspace}
\def\ccbar     {{\ensuremath{\cquark\cquarkbar}}\xspace}
\def\bquark    {{\ensuremath{\Pb}}\xspace}

\def\pion   {{\ensuremath{\Ppi}}\xspace}
\def\piz    {{\ensuremath{\pion^0}}\xspace}

\def\pip    {{\ensuremath{\pion^+}}\xspace}
\def\pim    {{\ensuremath{\pion^-}}\xspace}
\def\pipm   {{\ensuremath{\pion^\pm}}\xspace}

\def\kaon    {{\ensuremath{\PK}}\xspace}
  \def\Kbar    {{\kern 0.2em\overline{\kern -0.2em \PK}{}}\xspace}

\def\KorKbar    {\kern 0.18em\optbar{\kern -0.18em K}{}\xspace}

\def\Kp      {{\ensuremath{\kaon^+}}\xspace}
\def\Km      {{\ensuremath{\kaon^-}}\xspace}

  \def\Dbar    {{\kern 0.2em\overline{\kern -0.2em \PD}{}}\xspace}
\def\D       {{\ensuremath{\PD}}\xspace}

\def\DorDbar    {\kern 0.18em\optbar{\kern -0.18em D}{}\xspace}
\def\Dz      {{\ensuremath{\D^0}}\xspace}
\def\Dzb     {{\ensuremath{\Dbar{}^0}}\xspace}
\def\Dp      {{\ensuremath{\D^+}}\xspace}
\def\Dm      {{\ensuremath{\D^-}}\xspace}

\def\Dstarp  {{\ensuremath{\D^{*+}}}\xspace}

\def\Ds      {{\ensuremath{\D^+_\squark}}\xspace}
\def\Dsp     {{\ensuremath{\D^+_\squark}}\xspace}
\def\Dsm     {{\ensuremath{\D^-_\squark}}\xspace}
\def\Dspm    {{\ensuremath{\D^{\pm}_\squark}}\xspace}

\def\B       {{\ensuremath{\PB}}\xspace}
\def\Bbar    {{\ensuremath{\kern 0.18em\overline{\kern -0.18em \PB}{}}}\xspace}

\def\BorBbar    {\kern 0.18em\optbar{\kern -0.18em B}{}\xspace}

\def\Bu      {{\ensuremath{\B^+}}\xspace}

\def\Bp      {{\ensuremath{\Bu}}\xspace}

\def\Bd      {{\ensuremath{\B^0}}\xspace}
\def\Bs      {{\ensuremath{\B^0_\squark}}\xspace}
\def\Bsb     {{\ensuremath{\Bbar{}^0_\squark}}\xspace}

\def\jpsi     {{\ensuremath{{\PJ\mskip -3mu/\mskip -2mu\Ppsi\mskip 2mu}}}\xspace}

  \def\Y#1S{\ensuremath{\PUpsilon{(#1S)}}\xspace}% no space before {...}!

\def\proton      {{\ensuremath{\Pp}}\xspace}

\def\Lz          {{\ensuremath{\PLambda}}\xspace}
\def\Lbar        {{\ensuremath{\kern 0.1em\overline{\kern -0.1em\PLambda}}}\xspace}
\def\LorLbar    {\kern 0.18em\optbar{\kern -0.18em \PLambda}{}\xspace}

\def\Lb      {{\ensuremath{\Lz^0_\bquark}}\xspace}

\def\Lc      {{\ensuremath{\Lz^+_\cquark}}\xspace}

\newcommand{\decay}[2]{\ensuremath{#1\!\to #2}\xspace}         % {\Pa}{\Pb \Pc}

\def\to                 {\ensuremath{\rightarrow}\xspace}

\def\CP                {{\ensuremath{C\!P}}\xspace}

\def\AT#1     {\ensuremath{A_{\mathrm{T}}^{#1}}\xspace}           % 2

\def\C#1      {\ensuremath{\mathcal{C}_{#1}}\xspace}                       % 9
\def\Cp#1     {\ensuremath{\mathcal{C}_{#1}^{'}}\xspace}                    % 7
\def\Ceff#1   {\ensuremath{\mathcal{C}_{#1}^{\mathrm{(eff)}}}\xspace}        % 9  
\def\Cpeff#1  {\ensuremath{\mathcal{C}_{#1}^{'\mathrm{(eff)}}}\xspace}       % 7
\def\Ope#1    {\ensuremath{\mathcal{O}_{#1}}\xspace}                       % 2
\def\Opep#1   {\ensuremath{\mathcal{O}_{#1}^{'}}\xspace}                    % 7

\newcommand{\tev}{\ifthenelse{\boolean{inbibliography}}{\ensuremath{~T\kern -0.05em eV}}{\ensuremath{\mathrm{\,Te\kern -0.1em V}}}\xspace}
\newcommand{\gev}{\ensuremath{\mathrm{\,Ge\kern -0.1em V}}\xspace}
\newcommand{\mev}{\ensuremath{\mathrm{\,Me\kern -0.1em V}}\xspace}
\newcommand{\kev}{\ensuremath{\mathrm{\,ke\kern -0.1em V}}\xspace}
\newcommand{\ev}{\ensuremath{\mathrm{\,e\kern -0.1em V}}\xspace}
\newcommand{\gevc}{\ensuremath{{\mathrm{\,Ge\kern -0.1em V\!/}c}}\xspace}
\newcommand{\mevc}{\ensuremath{{\mathrm{\,Me\kern -0.1em V\!/}c}}\xspace}
\newcommand{\gevcc}{\ensuremath{{\mathrm{\,Ge\kern -0.1em V\!/}c^2}}\xspace}
\newcommand{\gevgevcccc}{\ensuremath{{\mathrm{\,Ge\kern -0.1em V^2\!/}c^4}}\xspace}
\newcommand{\mevcc}{\ensuremath{{\mathrm{\,Me\kern -0.1em V\!/}c^2}}\xspace}

\def\mum  {\ensuremath{{\,\upmu\mathrm{m}}}\xspace}

\def\mub{\ensuremath{{\mathrm{ \,\upmu b}}}\xspace}

\def\invfb   {\ensuremath{\mbox{\,fb}^{-1}}\xspace}

\newcommand{\stat}{\ensuremath{\mathrm{\,(stat)}}\xspace}
\newcommand{\syst}{\ensuremath{\mathrm{\,(syst)}}\xspace}

\def\gsim{{~\raise.15em\hbox{$>$}\kern-.85em
          \lower.35em\hbox{$\sim$}~}\xspace}
\def\lsim{{~\raise.15em\hbox{$<$}\kern-.85em
          \lower.35em\hbox{$\sim$}~}\xspace}

\def\sqs   {\ensuremath{\protect\sqrt{s}}\xspace}

\def\ptot       {\mbox{$p$}\xspace}
\def\pt         {\mbox{$p_{\mathrm{ T}}$}\xspace}

\def\evtgen     {\mbox{\textsc{EvtGen}}\xspace}

\def\geant      {\mbox{\textsc{Geant4}}\xspace}

\def\photos     {\mbox{\textsc{Photos}}\xspace}

\def\pythia     {\mbox{\textsc{Pythia}}\xspace}

\def\tell1  {TELL1\xspace}
\def\ukl1   {UKL1\xspace}

\newcommand{\eg}{\mbox{\itshape e.g.}\xspace}

\def\apVal         {-0.52} %
\def\apStat        {0.13} %
\def\apSyst        {0.10} %

\newcommand{\Araw}{\ensuremath{A_{\rm raw}}\xspace}

\newcommand{\AD}{\ensuremath{A_{\rm D}}\xspace}
\newcommand{\Apid}{\ensuremath{A_{\rm PID}}\xspace}
\newcommand{\Atrighard}{\ensuremath{A_{\rm trigger}^{\rm hardware}}\xspace}
\newcommand{\Atrigsoft}{\ensuremath{A_{\rm trigger}^{\rm software}}\xspace}
\newcommand{\AKK}{\ensuremath{A_{\rm track}^{KK}}\xspace}
\newcommand{\Api}{\ensuremath{A_{\rm track}^{\pi}}\xspace}

\newcommand{\APDs}{\ensuremath{A_{\rm P}(\Dsp)}\xspace}

\newcommand{\APB}{\ensuremath{A_{\rm P}(\B)}\xspace}

\newcommand{\fbkg}{\ensuremath{f_{\rm bkg}}\xspace}

\newcommand{\nKStar}{\ensuremath{K^*(892)^0}\xspace}

\def\y      {\ensuremath{y}\xspace}

\def\Dsstarp {{\ensuremath{\D_s^{*+}}}\xspace}

\newcommand{\TeV}{{\ensuremath{\mathrm{\,Te\kern -0.1em V}}}\xspace}
\usepackage{cite} % Allows for ranges in citations
\usepackage{mciteplus}

\begin{document}

%%%%%%%%%%%%%%%%%%%%%%%%%
%%%%% Title     %%%%%%%%%
%%%%%%%%%%%%%%%%%%%%%%%%%
\renewcommand{\thefootnote}{\fnsymbol{footnote}}
\setcounter{footnote}{1}

% %%%%%%% CHOOSE TITLE PAGE--------
% $Id: title-LHCb-PAPER.tex 122762 2018-08-14 11:38:27Z sklaver $
% ===============================================================================
% Purpose: LHCb-PAPER journal paper title page template
% Author: 
% Created on: 2010-09-25
% ===============================================================================

%%%%%%%%%%%%%%%%%%%%%%%%%
%%%%%  TITLE PAGE  %%%%%%
%%%%%%%%%%%%%%%%%%%%%%%%%
\begin{titlepage}
\pagenumbering{roman}

% Header ---------------------------------------------------
\vspace*{-1.5cm}
\centerline{\large EUROPEAN ORGANIZATION FOR NUCLEAR RESEARCH (CERN)}
\vspace*{1.5cm}
\noindent
\begin{tabular*}{\linewidth}{lc@{\extracolsep{\fill}}r@{\extracolsep{0pt}}}
\ifthenelse{\boolean{pdflatex}}% Logo format choice
{\vspace*{-1.5cm}\mbox{\!\!\!\includegraphics[width=.14\textwidth]{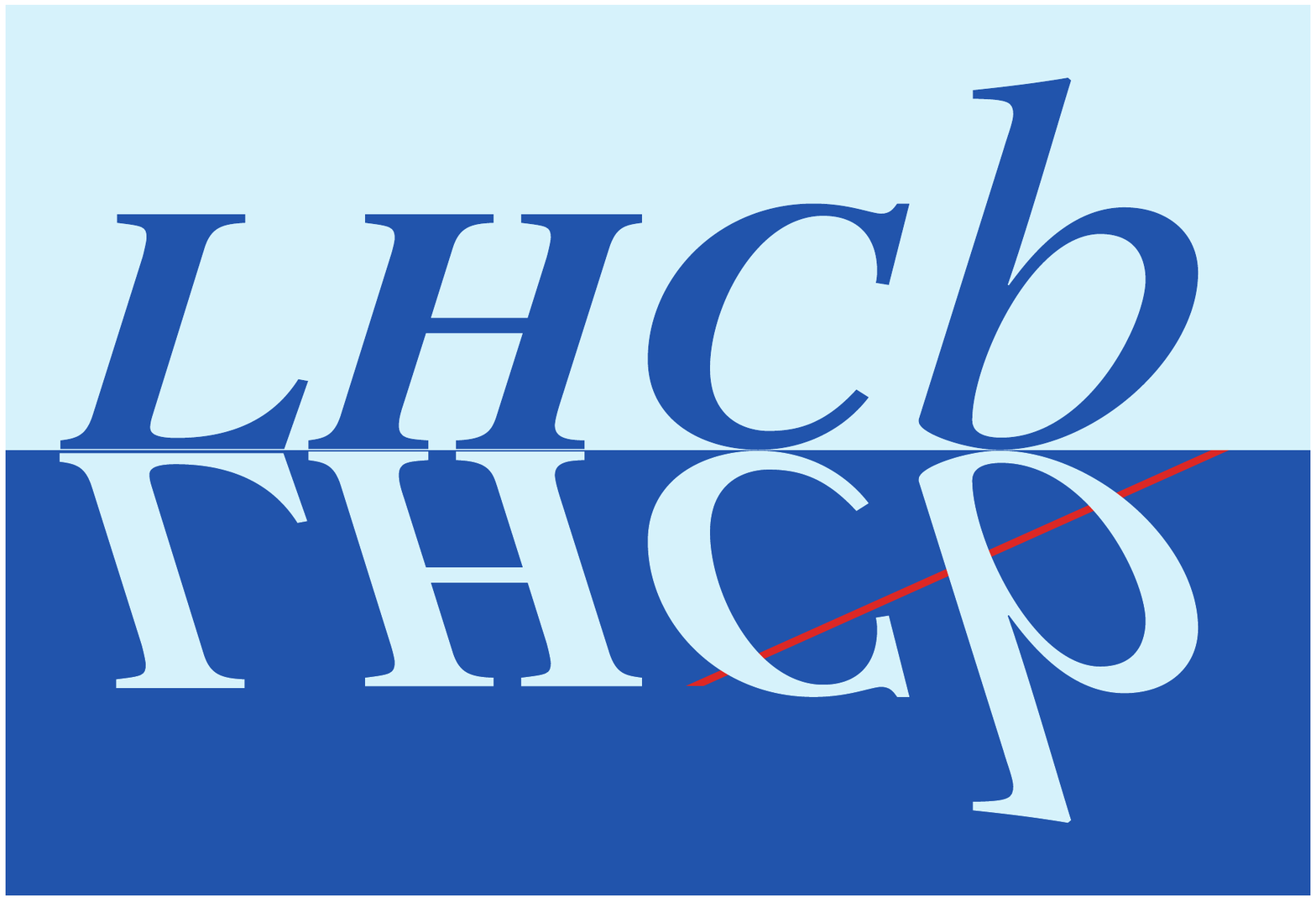}} & &}%
{\vspace*{-1.2cm}\mbox{\!\!\!\includegraphics[width=.12\textwidth]{lhcb-logo.eps}} & &}%
\\
 \vspace*{-2cm} \\
 & & CERN-EP-2018-073 \\  % ID 
 & & LHCb-PAPER-2018-010 \\  % ID 
 & & 14 August 2018 \\ % Date - Can also hardwire e.g.: 23 March 2010
\end{tabular*}

\vspace*{1.5cm}

% Title --------------------------------------------------
{\normalfont\bfseries\boldmath\huge
\begin{center}
  \papertitle 
\end{center}
}

\vspace*{1.0cm}

% Authors -------------------------------------------------
\begin{center}
\paperauthors\footnote{Authors are listed at the end of this paper.}
\end{center}

\vspace{\fill}

% Abstract -----------------------------------------------
\begin{abstract}
  \noindent
  The inclusive \Dspm production asymmetry is measured in $pp$ collisions
  collected by the \lhcb experiment
  at centre-of-mass energies of $\sqs=7$ and 8\tev. Promptly produced
  \Dspm mesons are used, which decay as $\Dspm\to\Pphi\pipm$, with
  $\Pphi\to\Kp\Km$. The measurement is performed in bins of
  transverse momentum, \pt, and rapidity, \y, covering the range $2.5<\pt<25.0
  \gevc$ and $2.0<y<4.5$. No kinematic dependence is observed. 
  Evidence of nonzero \Dspm production asymmetry is found with a significance
  of 3.3 standard deviations.
\end{abstract}

\vspace*{2.0cm}

\begin{center}
  Published in JHEP 08 (2018) 008 
\end{center}

\vspace{\fill}

{\footnotesize
\centerline{\copyright~\papercopyright. \href{\paperlicenceurl}{\paperlicence}.}}
\vspace*{2mm}

\end{titlepage}

%%%%%%%%%%%%%%%%%%%%%%%%%%%%%%%%
%%%%%  EOD OF TITLE PAGE  %%%%%%
%%%%%%%%%%%%%%%%%%%%%%%%%%%%%%%%

%  empty page follows the title page ----
\newpage
\setcounter{page}{2}
\mbox{~}

\cleardoublepage

% %%%%%%%%%%%%% ---------

\renewcommand{\thefootnote}{\arabic{footnote}}
\setcounter{footnote}{0}

%%%%%%%%%%%%%%%%%%%%%%%%%
%%%%% Main text %%%%%%%%%
%%%%%%%%%%%%%%%%%%%%%%%%%

\pagestyle{plain} % restore page numbers for the main text
\setcounter{page}{1}
\pagenumbering{arabic}

% $Id: introduction.tex 122762 2018-08-14 11:38:27Z sklaver $

\section{Introduction}
\label{sec:Introduction}

At \lhc energies, \ccbar quark pairs are copiously produced in $pp$ collisions 
with a total cross-section at a centre-of-mass energy of 7\tev of 
$\sigma_{\ccbar} = 1419 \pm 136$\mub\cite{LHCB-PAPER-2012-041}.
The subsequent hadronisation process can
introduce a charge asymmetry in the production of charm hadrons. This asymmetry
is influenced by the valence quarks of the colliding protons, which results in 
a preference for the \cquarkbar quark to form a meson, \eg a \Dm or
\Dzb meson. A \cquark quark, on the other hand, can form charm baryons, \eg a
\Lc baryon, with the proton's valence quarks. This difference in hadronisation
gives rise to different kinematic distributions between charge-conjugated charm
hadrons, and therefore results in a charge asymmetry.

The \Dsp meson does not contain any of the proton's valence quarks, which means
that the aforementioned processes can contribute only indirectly to a production
asymmetry. The \Dsp production asymmetry is defined as
\begin{equation}
\APDs = \frac{\sigma(\Dsp) - \sigma(\Dsm)}{\sigma(\Dsp) + \sigma(\Dsm)} \, ,
\label{eq:ApDs}
\end{equation}
where $\sigma(\Dspm)$ is the inclusive prompt production cross-section. It is
difficult to make accurate predictions of the \Dsp production asymmetry due to
the nonperturbative nature of the hadronisation process. Nonetheless, the Lund
string fragmentation model~\cite{Andersson:1983ia}, implemented in
\pythia\cite{Sjostrand:2007gs, *Sjostrand:2006za}, describes
hadronisation that can give rise to production asymmetries for heavy
flavours~\cite{Norrbin:1999by, Norrbin:2000jy, Norrbin:2000zc}. This model
predicts that production asymmetries can be dependent on kinematics due to
interactions with the beam remnants. A measurement of the \Dsp production asymmetry 
can be used to test nonperturbative QCD models and is an essential input for 
measurements of direct \CP violation in the decays of \Ds mesons in \lhcb. 

This paper presents a measurement of the \Dsp production asymmetry in $pp$
collisions using two data sets corresponding to integrated luminosities of
$1.0\invfb$ and $2.0\invfb$, recorded by the \lhcb detector at centre-of-mass energies
of 7 and 8\tev in 2011 and 2012, respectively. An inclusive sample of promptly
produced \Dsp mesons in the decay mode \decay{\Dsp}{\Pphi\pip} is used, 
where \decay{\Pphi}{\Kp\Km}. 
This sample
includes excited states that decay to \Dsp mesons, such as \Dsstarp mesons which
decay to \Dsp\g or \Dsp\piz. 
The inclusion of charge-conjugate processes is
implied throughout this paper, except in the definition of the asymmetries.

The \Dsp production
asymmetries derived from \pythia are compared to the results obtained in this paper. A
previous measurement by the \lhcb collaboration~\cite{LHCb-PAPER-2012-009} with the 7\tev data set
indicated a small excess of \Dsm over \Dsp mesons, resulting in a negative value
for the production asymmetry. This paper, with improvements in the detector calibration, 
supersedes the previous measurement and includes the 8\tev data set.

\section{Detector and simulation}
\label{sec:Detector}

The \lhcb detector~\cite{Alves:2008zz,LHCb-DP-2014-002} is a single-arm forward
spectrometer covering the \mbox{pseudorapidity} range $2<\eta <5$,
designed for the study of particles containing \bquark or \cquark
quarks. The detector includes a high-precision tracking system
consisting of a silicon-strip vertex detector surrounding the $pp$
interaction region,
a large-area silicon-strip detector located
upstream of a dipole magnet with a bending power of about
$4{\mathrm{\,Tm}}$, and three stations of silicon-strip detectors and straw
drift tubes
placed downstream of the magnet.
The polarity of the dipole magnet is reversed periodically throughout 
data taking and the corresponding data sets (referred to as \MagUp and \MagDown) are approximately equal in size.
The tracking system provides a measurement of momentum, \ptot, of charged particles with
a relative uncertainty that varies from 0.5\% at low momentum to 1.0\% at 200\gevc.
The minimum distance of a track to a primary vertex (PV), the impact parameter (IP), 
is measured with a resolution of $(15+29/\pt)\mum$,
where \pt is the component of the momentum transverse to the beam, in\,\gevc.
Different types of charged hadrons are distinguished using information
from two ring-imaging Cherenkov detectors.
Photons, electrons and hadrons are identified by a calorimeter system consisting of
scintillating-pad and preshower detectors, an electromagnetic
calorimeter and a hadronic calorimeter. Muons are identified by a
system composed of alternating layers of iron and multiwire
proportional chambers.
The online event selection is performed by a trigger~\cite{LHCb-DP-2012-004}, 
which consists of a hardware stage, based on information from the calorimeter and muon
systems, followed by a software stage, which applies a full event
reconstruction.

In the simulation, which is used for comparing the production asymmetry results, 
$pp$ collisions are generated using
\pythia~\cite{Sjostrand:2007gs,*Sjostrand:2006za}, which has implemented the Lund
string fragmentation model~\cite{Andersson:1983ia},
with a specific \lhcb configuration~\cite{LHCb-PROC-2010-056}.
Decays of hadronic particles are described by \evtgen~\cite{Lange:2001uf},
in which final-state radiation is generated using \photos~\cite{Golonka:2005pn}.
The interaction of the generated particles with the detector, and its response,
are implemented using the \geant toolkit~\cite{Allison:2006ve, *Agostinelli:2002hh}
as described in Ref.~\cite{LHCb-PROC-2011-006}.

\section{Data selection}
\label{sec:dataset}

Signal candidates are selected by the requirements made in the
trigger and in the offline selection.
At the hardware trigger stage, events are required to have 
a muon with high transverse momentum or a hadron, photon or electron with 
high transverse energy deposited in the calorimeters. 
The software trigger requires at least one charged particle
that has $\pt > 1.7\gevc$ at 7\tev or $\pt > 1.6\gevc$ at 8\tev,
and is inconsistent with originating from any PV.
Subsequently, three well reconstructed tracks
are required to originate from
a common vertex with a significant displacement from any PV. 
Additional requirements are made to select three-prong decays with 
an invariant mass close to that of the \Dsp meson. The 
reconstructed \Ds meson must have $\pt>2.5\gevc$.

In the offline selection, trigger decisions are associated with
reconstructed tracks or energy deposits. Requirements 
can therefore be made on whether the trigger 
decision was due to the signal candidate, other particles
in the event, or a combination of both.
For the hardware trigger stage, a positive trigger decision
is required to be caused by a particle that is distinct from any
of the final-state particles that compose the \Dsp candidate.
This requirement is independent of whether or not the signal 
candidate itself also caused a positive trigger decision, and 
is therefore referred to as triggered independently of signal 
(TIS)~\cite{LHCb-DP-2012-004}. For the software trigger stage, 
the positive trigger decision is required to be associated with 
the final-state particles of the \Dsp candidate. This is called 
triggered on signal (TOS)~\cite{LHCb-DP-2012-004}.

The three tracks from the final-state particles 
are required to not point back to any PV. To reduce candidates
from \bquark-hadron decays, the \Dsp 
candidate itself must point to a PV. Its decay vertex is required 
to have a good quality and to be significantly displaced from any PV.
To ensure a good overlap with the additional samples used for 
calibration purposes, $\ptot > 5.0$ (3.0)\gevc and $\pt > 400$
(300)\mevc are required for the pions (kaons). 
Background due to random combinations of tracks 
is suppressed by requiring the sum of the \pt of the final-state 
tracks to be larger than 2.8\gevc.

Kaon and pion mass assignments to the particle tracks are
based on the particle identification 
(PID) information obtained primarily from the Cherenkov detectors.
The invariant mass of the kaon pair is 
required to be within 20\mevcc of the known \Pphi mass~\cite{PDG2017}.
The mass of the \Ds candidate is selected to be 
between 1900 and 2035\mevcc. Additional PID and mass requirements 
are applied to suppress two particular sources of background.
The first comes from \decay{\Lc}{\proton \Km \pip} decays, where the 
proton is misidentified as a kaon. The second are \decay{\Dp}{\Km\pip\pip}
decays, where one of the pions is misidentified as a kaon. Both of these are 
suppressed by applying tighter PID requirements in a small window of 
invariant mass of the corresponding particle combination around the 
known \Lc and \Dp masses. The remaining contribution from misidentified 
\Lc and \Dp decays is negligibly small.

After the full selection, $2.9\times10^6$ and $9.1\times10^6$ \Ds candidates
are selected in the 7\tev and 8\tev data sets, respectively, with
a signal purity of 97\%. The increase for the 8\tev data set is not only due to
a higher integrated luminosity, but also to improvements in the trigger. 
The fraction of events with more than one candidate, which are not removed
in this analysis, is only $2\times10^{-4}$, resulting in a negligible bias in the final asymmetry.
The two data sets with opposite magnetic fields are analysed separately.

\section{Analysis method}
\label{sec:analysis}

The raw asymmetry is defined as the difference between the observed 
numbers, $N(\Dspm)$, of \Dsp and \Dsm mesons
\begin{align}
  \Araw = \frac{N(\Dsp) - N(\Dsm)}{N(\Dsp) + N(\Dsm)} \, .
  \label{eq:Araw}
\end{align}
This asymmetry
must be corrected for contributions from \Dsp mesons originating
from \bquark-hadron decays, and for detection asymmetries, \AD. 
The production asymmetry, assuming the \CP asymmetry in Cabibbo-favoured 
\Dsp decays to be negligible at the precision of this measurement,
 is determined as
\begin{align}
  \APDs = \frac{1}{1-\fbkg} (\Araw - \AD - \fbkg \APB) \, ,
  \label{eq:ApDs2}
\end{align}
where $\fbkg$ is the fraction of \Dsp mesons that originate from
\bquark-hadron decays and \APB the production asymmetry of these \bquark
hadrons. 

\begin{figure}\centering
\includegraphics[width=0.49\textwidth]{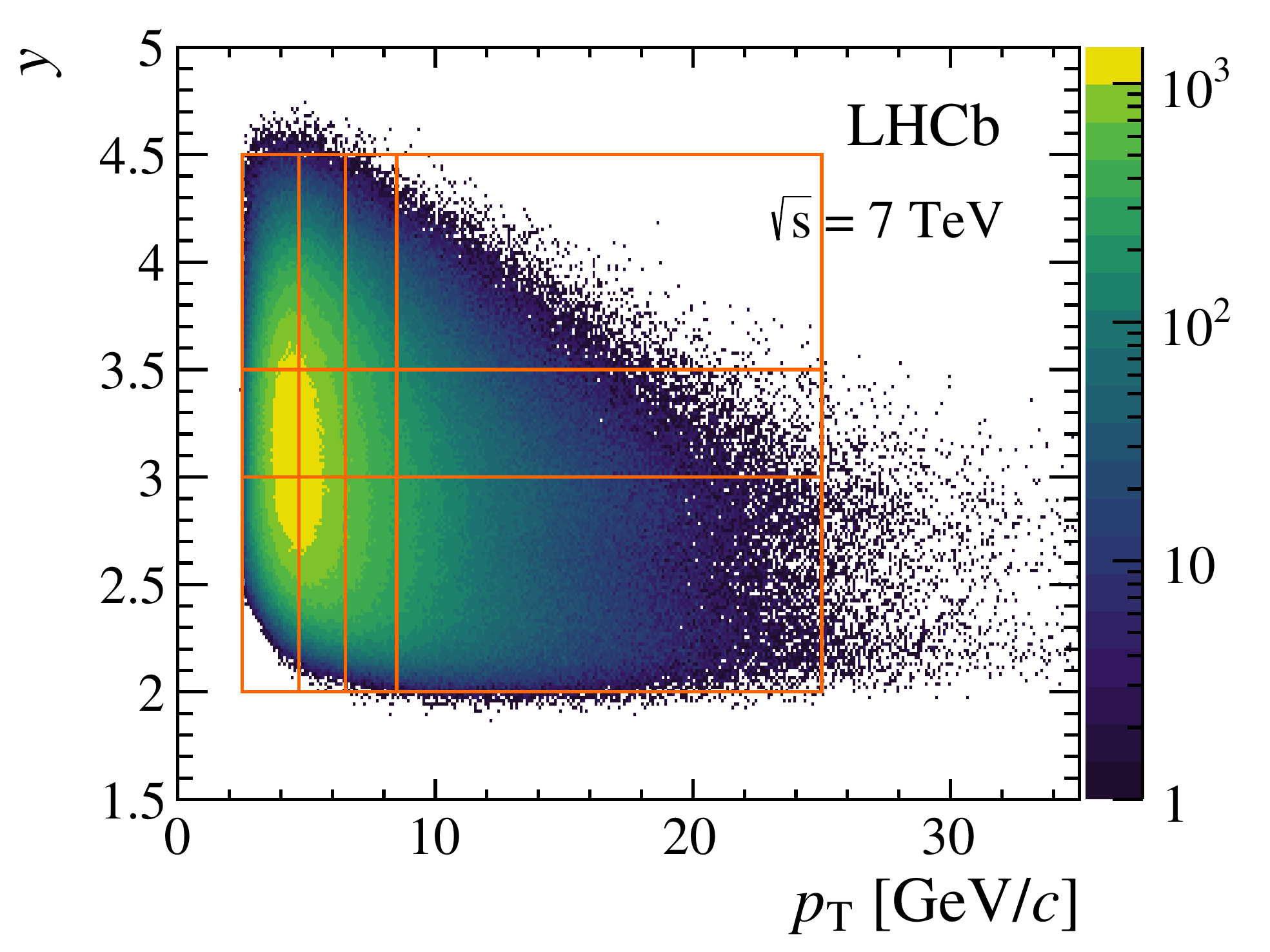}
\includegraphics[width=0.49\textwidth]{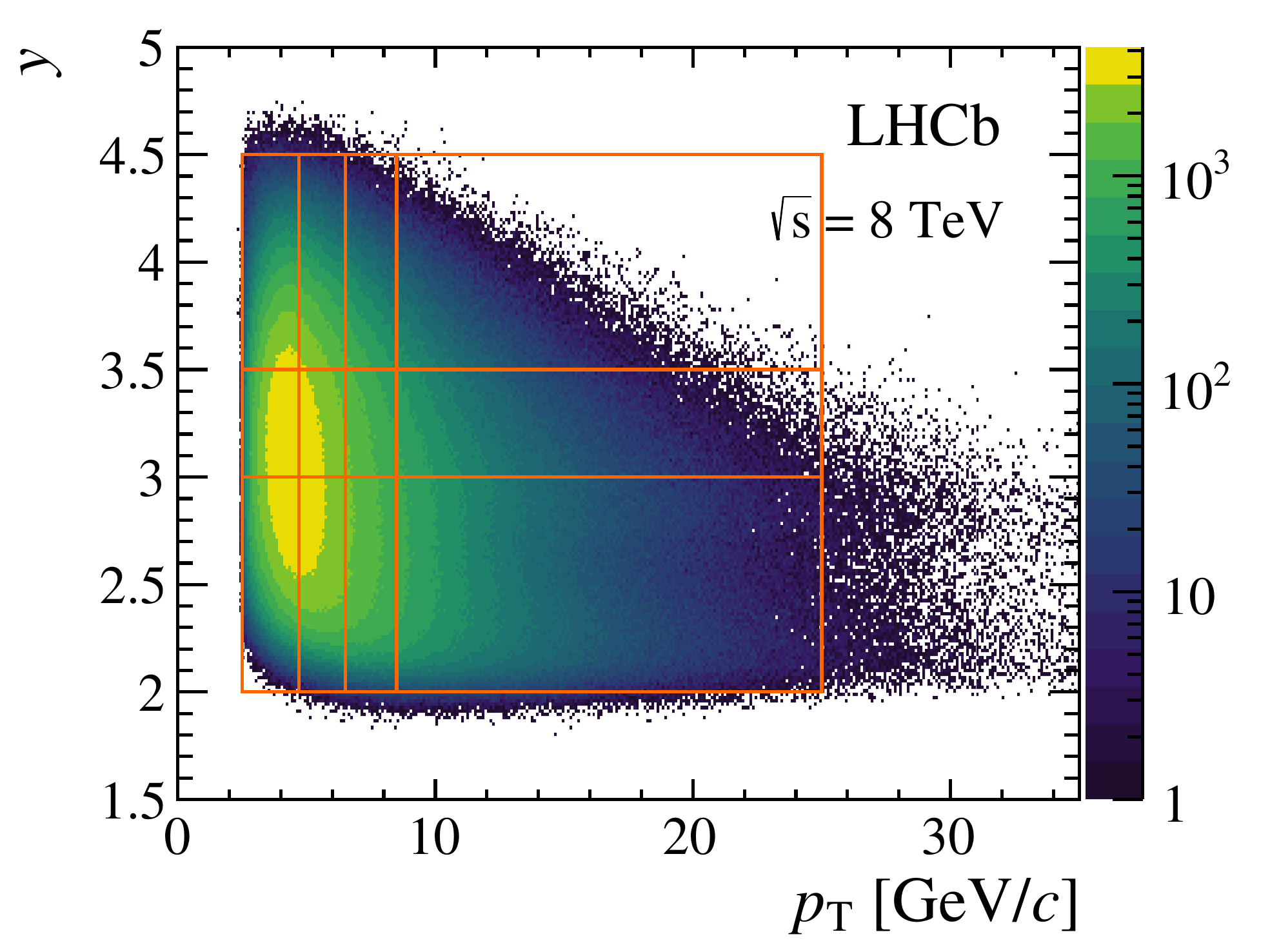}
\caption{\label{fig:dsbinning} Distribution of \Dsp candidates for the $\sqs=7$ and 8 TeV data sets as a 
function of \pt and \y. The binning scheme used for the \APDs measurement is overlaid.}
\end{figure}

Since the production asymmetry may depend on the kinematics of the \Dsp meson, 
the measurement is performed in two-dimensional bins of \pt and rapidity, \y. 
Four bins in \pt and three bins in \y are chosen as follows
\begin{align*}
  \pt \; [\!\gevc] : &\;[2.5 , 4.7]\, ;\, [4.7 , 6.5]\,;\, [6.5 , 8.5]\,;\, [8.5 , 25.0] \, ,\\
  \y: &\; [2.0 , 3.0]\,;\, [3.0 , 3.5]\,;\, [3.5 , 4.5] \, ,
\end{align*}
where the rapidity of the \Dsp meson is defined as
\begin{align}
  y = \frac{1}{2} \ln{\left(\frac{E+p_z c}{E-p_z c}\right) } \, .
  \label{eq:rapidity}
\end{align}
Here, $E$ is the energy of the \Dsp meson and $p_z$ the component of its momentum
along the beam direction. 
This binning scheme is chosen such that the bins are roughly equally populated and is 
the same as that used for the previous \APDs measurement~\cite{LHCb-PAPER-2012-009}, 
except that the lowest \pt bin is now split into two.
The two-dimensional distribution in \pt and \y
of \Dsp candidates is shown in Fig.~\ref{fig:dsbinning} along with the binning
scheme.

\subsection{Measurement of raw asymmetries}

\begin{figure}\centering
\includegraphics[trim={0.25cm 0.3cm 0cm 0cm},clip,width=0.48\textwidth]{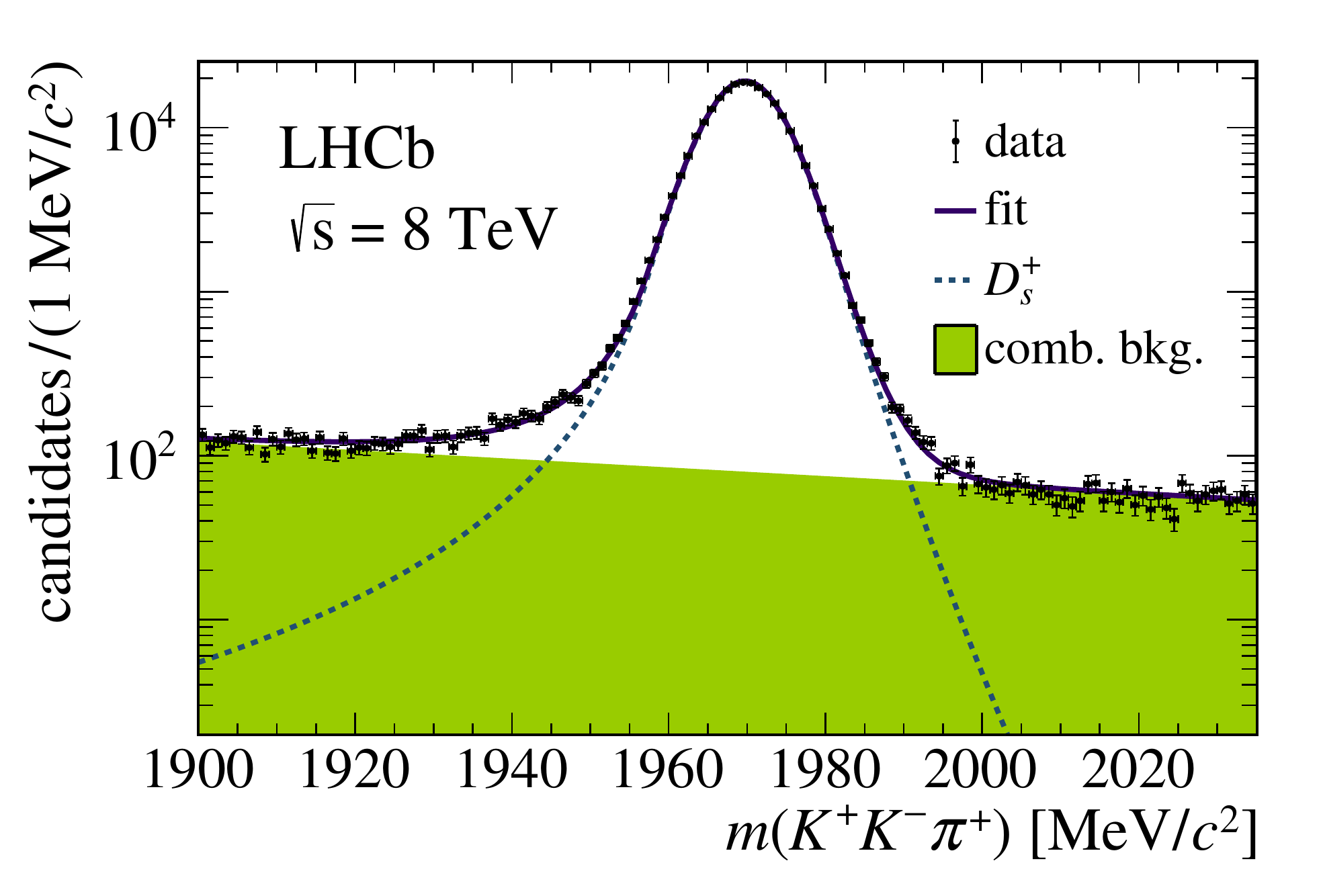}
\hspace{0.01\textwidth}
\includegraphics[trim={0.25cm 0.3cm 0cm 0cm},clip,width=0.48\textwidth]{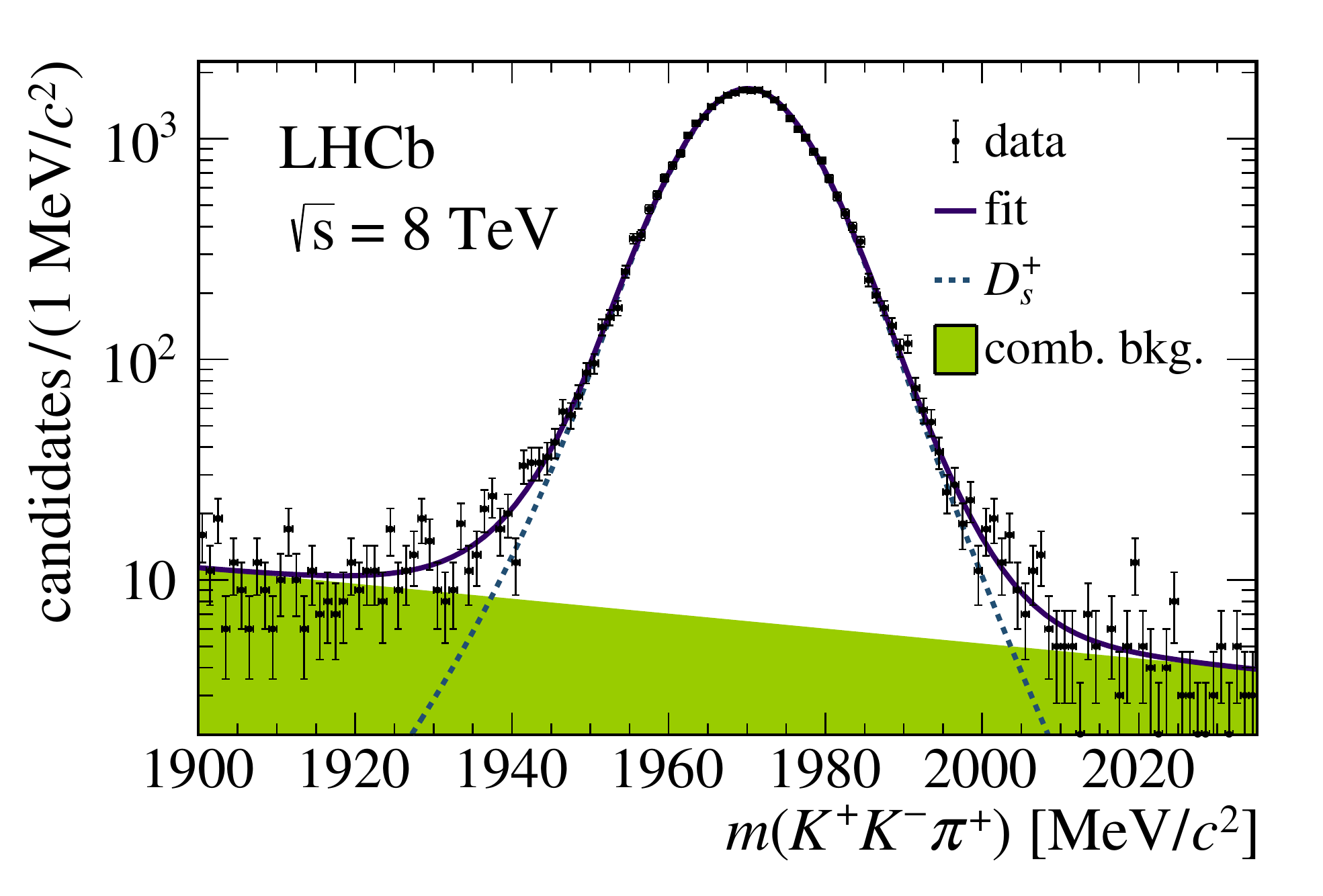}
\caption{\label{fig:Araw}Invariant mass distribution of the \Dsp candidates 
in the kinematic range (left) $2.0<y<3.0$ and $2.5<\pt<4.7\gevc$, and (right) $3.5<y<4.5$ and $8.5<\pt<25.0\gevc$ for the $\sqs=8$ TeV data set 
recorded with \MagDown. Also shown is the result of the fit, indicating the signal and combinatorial background.}
\end{figure}

The signal yields and asymmetries are obtained from binned maximum-likelihood
fits to the \Dsp mass distributions in the twelve kinematic bins, separately for
the two data-taking periods and the two magnet polarities. The signal component is
modelled with a Hypatia function with tails on both sides~\cite{Santos:2013gra}, and the
combinatorial background, from random combinations of tracks,
with an exponential function. The parameters describing
the tails of the Hypatia function are determined by fits to the \Dsp mass distributions
that are performed in each kinematic bin, in which the data sets from 7 and 8\tev 
and both magnet polarities are combined. These
parameters are then kept fixed in the fits to obtain the raw asymmetries. The
raw asymmetries in each kinematic bin are obtained from simultaneous fits to the \Dsp and \Dsm mass
distributions in which all free parameters are shared, except for the yields, the
mean mass of the signal component, and the background parameters. The mean mass can be different as the 
momentum reconstruction may have different biases for positive and negative tracks.
The variation of the background parameters is needed to account for potential 
asymmetries in the background.
Two example fits are shown in Fig.~\ref{fig:Araw}.

A systematic uncertainty is assigned for the effect of fixing the tail parameters
by varying their values and reassessing the raw asymmetries. 
In addition, a possible bias from the fit model is studied by generating 
invariant mass distributions with the signal component described by a double
Gaussian function with power-law tails on both sides, which are subsequently 
fitted using the default Hypatia function. The differences in the raw asymmetry
for both studies are assigned as a systematic uncertainty. This is small because 
the low amount of combinatorial background allows for little bias from the fit model. 

\subsection[Contribution from b-hadron decays]{{\boldmath Contribution from \bquark-hadron decays}}

An estimate of the fraction of \Dsp candidates from \bquark-hadron (\Bp, \Bd, \Bs, \Lb) 
decays is performed using a combination of simulation, known
cross-sections~\cite{LHCb-PAPER-2013-004, LHCb-PAPER-2014-004} and known
branching fractions~\cite{PDG2017}.
Simulation samples are used to determine the reconstruction and selection efficiencies 
relative to those for the signal decay. The fraction of \Dsp from \bquark-hadron decays is estimated to
be $\fbkg = (4.12\pm1.23) \%$, where the uncertainty includes contributions from the experimental
input and the simulation.

The production asymmetries for \bquark hadrons are taken from measurements from the \lhcb 
collaboration~\cite{LHCb-PAPER-2016-054,LHCB-PAPER-2016-062,LHCb-PAPER-2014-053}. 
The \Bs production asymmetry is diluted due to the fact that, before it decays, a \Bs meson oscillates with high frequency to a \Bsb, and vice versa. Hence, its contribution is assumed to be zero. 
Wherever multiple \lhcb measurements are available using different
decay channels, their results are combined in a weighted average. 
The contribution of the background asymmetry to \APDs, as defined in Eq.~\ref{eq:ApDs2}, is found to be
\begin{align*}
  \fbkg\APB = (0.3 \pm 1.0) \times 10^{-4} \ \, \rm{at} \  7\tev , \\
  \fbkg\APB = (1.7 \pm 0.8) \times 10^{-4} \ \, \rm{at} \ 8\tev  ,
\end{align*}
which is very small compared to the experimental precision of the measurement. The dilution from \fbkg 
in the denominator of Eq.~\ref{eq:ApDs2} gives a small correction to \Araw.

\section{Detection asymmetries}
\label{sec:detection_asymmetries}

Detection asymmetries are caused by the differences in reconstruction
efficiencies between \Dsp and \Dsm mesons and originate from the various
stages in the reconstruction process. Since these asymmetries are small, they factorise and can be added 
up as
\begin{align}
  \AD =  \Api + \AKK + \Apid + \Atrigsoft + \Atrighard  .
  \label{eq:AD}
\end{align}
Here, \Api and \AKK are the tracking asymmetries of the pion and the 
kaon pair, respectively. The asymmetry originating from the PID
requirements in the selection is denoted by \Apid. Lastly,
asymmetries arising from the trigger are split between the hardware
and software components of the trigger as \Atrighard and \Atrigsoft.
All detection asymmetries are determined for and applied to each bin using data-driven methods
described below, and corrected by simulations wherever necessary. 
Values of the detection asymmetries for the 7 and 8\tev data sets, 
determined on the data combined from all kinematic bins, are listed in Table~\ref{tab:DsSyst}.
In this paper, the statistical uncertainties from the detection asymmetries, 
obtained from control samples, are included in the total statistical uncertainty of the measurement.

\begin{table}
\setlength{\tabcolsep}{1em}
  \centering
  \caption{Raw and detection asymmetries in percent, for the 7 and 8 TeV data sets.
  The detection asymmetries are determined on the data combined from all kinematic bins.  
  The first uncertainty is statistical, the second systematic.}
  \label{tab:DsSyst} 
    \bgroup
\def\arraystretch{1.2}
  \begin{tabular}{l *{2}{r@{ $\pm$ }l@{ $\pm$ }l } l }
  \toprule
  source         & \multicolumn{3}{c}{$\sqs=7\tev$} & \multicolumn{3}{c}{$\sqs=8\tev$} \\
  \midrule
  \Araw 		& $-0.431$ & $0.061$ & $0.006$ 	&  $-0.492$ & $0.034$ & $0.006$ \\
  \Api 			& $ 0.093$ & $0.096$ & $0.048$ 	&  $-0.026$ & $0.068$ & $0.048$ \\
  \AKK  		& $ 0.000$ & $0.000$ & $0.030$ 	&  $ 0.000$ & $0.000$ & $0.030$ \\
  \Apid 		& $-0.018$ & $0.008$ & $0.012$ 	&  $ 0.008$ & $0.005$ & $0.012$ \\
  \Atrighard  	& $ 0.139$ & $0.229$ & $0.066$ 	&  $-0.060$ & $0.115$ & $0.066$ \\
  \Atrigsoft	& $-0.005$ & $0.018$ & $0.033$ 	&  $ 0.026$ & $0.011$ & $0.033$ \\
  \midrule
  \APDs         & $-0.671$ & $0.267$ & $0.095$  &  $-0.477$ & $0.145$ & $0.095$  \\
  \bottomrule
 \end{tabular}
\egroup
\end{table}

\subsection{Tracking asymmetries}

When the kaons originate from the \Pphi resonance, there can
be no detection asymmetry from the kaon pair. Only the small
fraction of kaons coming from the nonresonant decays included in the selection 
can introduce a detection asymmetry, and only when
the kinematic distributions of the two oppositely charged kaons
are different. In general,
the reconstruction efficiency of kaons suffers from a sizeable 
difference between the interaction cross-sections of \Kp and \Km mesons with the
detector material, which depends on the kaon momentum. 
For the pair of kaons, however, these differences largely cancel, since the 
momentum distributions of the positively and negatively charged 
kaons are very similar.
An upper limit of $3\times10^{-4}$ is set on their contribution, 
based on their kinematic overlap and the maximum kaon detection asymmetry 
as measured with calibration data~\cite{LHCb-PAPER-2014-013,LHCb-PAPER-2016-013}.

The pion tracking asymmetry is determined using two different methods, 
analogously to Ref.~\cite{LHCb-PAPER-2016-013}. The first uses muons from
partially reconstructed \decay{\jpsi}{\mup\mun} decays, as described in Ref.~\cite{LHCb-DP-2013-002}.
The second method uses partially reconstructed 
\decay{\Dstarp}{\Dz\pip} decays with \decay{\Dz}{\Km\pip\pim\pip}, 
where one of the pions from the \Dz decay does not need to be 
reconstructed~\cite{LHCb-PAPER-2012-009}. Both methods have limitations:
the former because it does not probe the full detector
acceptance or the effect of the hadronic interaction of the pion 
with the detector material, the latter because it is limited to pions 
with momenta
below 100\gevc. The limitations on the \decay{\jpsi}{\mup\mun} method
are assessed and corrected using simulation. After these corrections, 
the two methods are in good agreement and
the final value of \Api is determined by the weighted average of the two methods. 
For pions with $\ptot>100\gevc$, \Api is determined solely using the
\decay{\jpsi}{\mup\mun} method combined with the above-mentioned corrections.

\subsection{Particle identification asymmetries}

The asymmetry induced by the PID requirements, \Apid, is determined using large
samples of \decay{\Dstarp}{\Dz\pip} decays, with
\decay{\Dz}{\Km\pip}~\cite{Anderlini:2202412}. The \Dstarp charge identifies
which of the two particles is the kaon and which the pion in the \Dz decay without the use of PID
requirements. These unbiased samples are then used to determine the PID
efficiencies and corresponding charge asymmetries.

\subsection{Trigger asymmetries}

The efficiencies of the hardware and software triggers are studied using the signal sample
of prompt \decay{\Dsp}{\Kp\Km\pip} decays. For the hardware trigger, the TIS
asymmetry is determined with respect to decays that are TOS as well as TIS. This is done by
evaluating the TIS asymmetry separately for candidates that are TOS,
triggered by the \Kp track, and candidates that are TOS, triggered by the \Km track, and then
averaging the asymmetry. Due to possible correlations between the
signal decay and the rest of the event, this method is biased. 
In addition, as a result of the coarse transverse segmentation of the
hadronic calorimeter, the energy deposited by other particles in the event can
increase the energy that is measured and associated to the signal tracks of TOS events. 
This further increases the bias of the measured TIS asymmetry.
To assess the systematic uncertainty from this
method, a much larger sample of \decay{\Dp}{\Km\pip\pip} decays is studied.
In this sample the TIS asymmetries are determined using candidates that are TOS,
triggered by one, two or all three of the final-state particles.
The difference in the asymmetry resulting from these variations is
assigned as the systematic uncertainty.

The asymmetry due to the software trigger is assessed by the 
TOS efficiency of a single track from the \Kp\Km\pip final state
in events that have been triggered by one of the other tracks. The individual 
efficiencies are determined in bins of transverse momentum and 
pseudorapidity, and are then combined to obtain the overall asymmetry
introduced by the software trigger selection. The systematic uncertainty
is determined by studying the effect of the 
difference between the online and offline determination of the 
transverse momentum, and by determining the effect of the binning scheme.

\subsection{Systematic uncertainties}
In addition to the systematic uncertainties discussed above, all detection asymmetries are determined in bins 
of kinematic variables of final-state particles, for example \pt, $\eta$. The 
limitations of the binning schemes are evaluated by changing to different binning 
schemes, \eg \ptot, $\eta$. The systematic uncertainties are fully correlated 
between the 7\tev and 8\tev data sets. An overview of the systematic and statistical 
uncertainties for both data sets from the various sources of detection asymmetries 
is shown in Table~\ref{tab:DsSyst}.

\section{Results}
\label{sec:results_systematics}

The values of the \Dsp production asymmetry obtained using the \MagUp and \MagDown data
sets separately are compatible with each other in each kinematic bin within two standard deviations, 
as illustrated in Fig.~\ref{fig:ApDsResultsPerBin_updown} in the Appendix. The two magnet
polarities are combined using the arithmetic mean to ensure that any residual
magnet-polarity-dependent detection asymmetry cancels. Due to the small
difference in centre-of-mass energy and since the observed production asymmetries are
statistically compatible, the 7\tev and 8\tev data sets are combined in a weighted average,
maximising the statistical precision. The resulting production asymmetries in each kinematic bin are
presented in Table~\ref{table:results}. The results for both centre-of-mass energies 
separately are provided in Tables~\ref{table:results2011} and~\ref{table:results2012} in the Appendix. 

Since no kinematic dependence is observed, the data from all kinematic bins are combined
and the full procedure is repeated to obtain values equivalent to a weighted average based on
the signal yields, taking into account the correlations from the calibration samples between the kinematic bins. 
These are shown for the 7 and 8\tev data sets separately in Table~\ref{tab:DsSyst}. 
Taking the weighted average of these two results with the systematic uncertainties as fully correlated, the combined value is
\begin{align*}
\APDs = (\apVal\pm\apStat\stat\pm\apSyst\syst)\% \, ,
\end{align*}
corresponding to a deviation of 3.3 $\sigma$ from the hypothesis of no production asymmetry.

\begin{table}[b]
\caption{\label{table:results} Values of the $\Dsp$ production asymmetry 
in percent, including, respectively, the statistical and systematic uncertainties
for each of the \Dsp kinematic bins using the combined $\sqs=7$ and $8\TeV$ 
data sets. The statistical and systematic uncertainties include the corresponding
contributions from the detection asymmetries, and are therefore correlated between the bins.}
\begin{center}
\bgroup
\def\arraystretch{1.2}
\begin{tabular}{l @{\hskip 1.75em} *{3}{r@{ $\pm$ }l@{ $\pm$ }l } }
\toprule
& \multicolumn{9}{c}{\y} \\
\cmidrule(l{-0.5em}){2-10}
\pt [\gevc]\hskip 1.25em & \multicolumn{3}{c}{$2.0-3.0$} & \multicolumn{3}{c}{$3.0-3.5$} & \multicolumn{3}{c}{$3.5-4.5$} \\
\cmidrule[0.5pt](r{1.2em}){1-1} 
\cmidrule[0.5pt](l{-0.5em}){2-10} 
$2.5-4.7$ & $-0.63$ & $0.34$ & $0.32$ & $-0.66$ & $0.31$ & $0.13$ & $-0.65$ & $0.33$ & $0.14$  \\
$4.7-6.5$ & $-0.68$ & $0.25$ & $0.27$ & $-0.06$ & $0.26$ & $0.10$ & $-0.72$ & $0.26$ & $0.13$  \\
$6.5-8.5$ & $-0.55$ & $0.22$ & $0.06$ & $-0.57$ & $0.26$ & $0.10$ & $-0.48$ & $0.30$ & $0.17$  \\
$8.5-25.0$ & $-0.40$ & $0.15$ & $0.08$ & $-0.24$ & $0.22$ & $0.10$ & $-0.86$ & $0.33$ & $0.09$ \\
\bottomrule
\end{tabular}
\egroup
\end{center}
\end{table}

The results presented here are in agreement with the 
previous measurement of the \Dsp production asymmetry~\cite{LHCb-PAPER-2012-009}, 
obtained using only the 7\tev data set.
A cross-check is performed by measuring \APDs in two other disjoint regions
in the \decay{\Dsp}{\Kp\Km\pip} Dalitz plot, analogous to those defined in 
Ref.~\cite{LHCb-PAPER-2016-013}. These are the region including the
\nKStar resonance, and the remaining nonresonant region. The \APDs 
measurements in the three regions are in good agreement in all kinematic bins. 
However, these regions are not included in the measurement of \APDs, since it
was found that including them slightly increases the uncertainty on 
the measurement due to the larger systematic effects from the detection asymmetries. 

\subsection{Comparison with P{\small YTHIA} predictions}

\begin{figure}[t]
\centering
\includegraphics[width=0.49\textwidth]{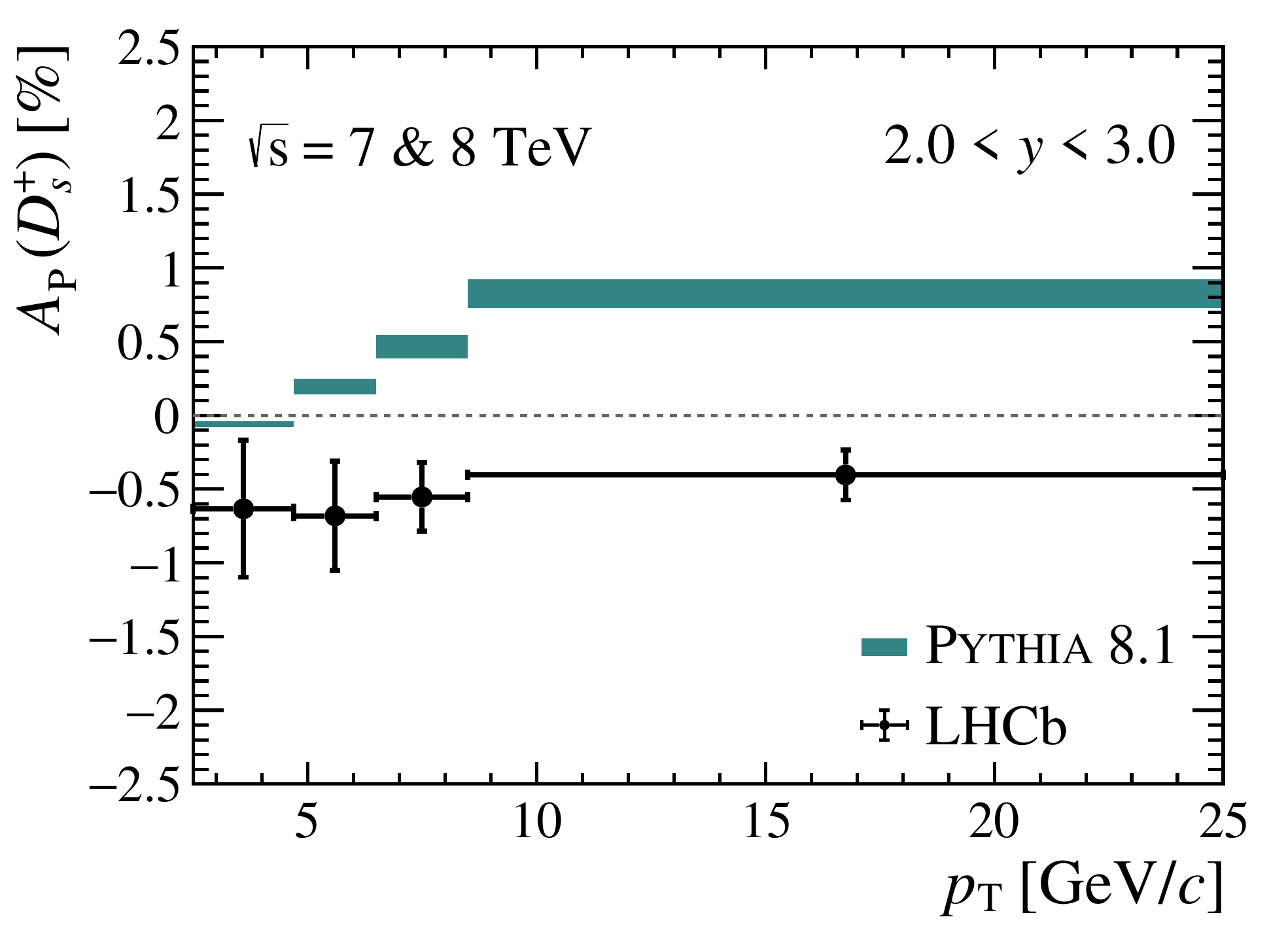}
\includegraphics[width=0.49\textwidth]{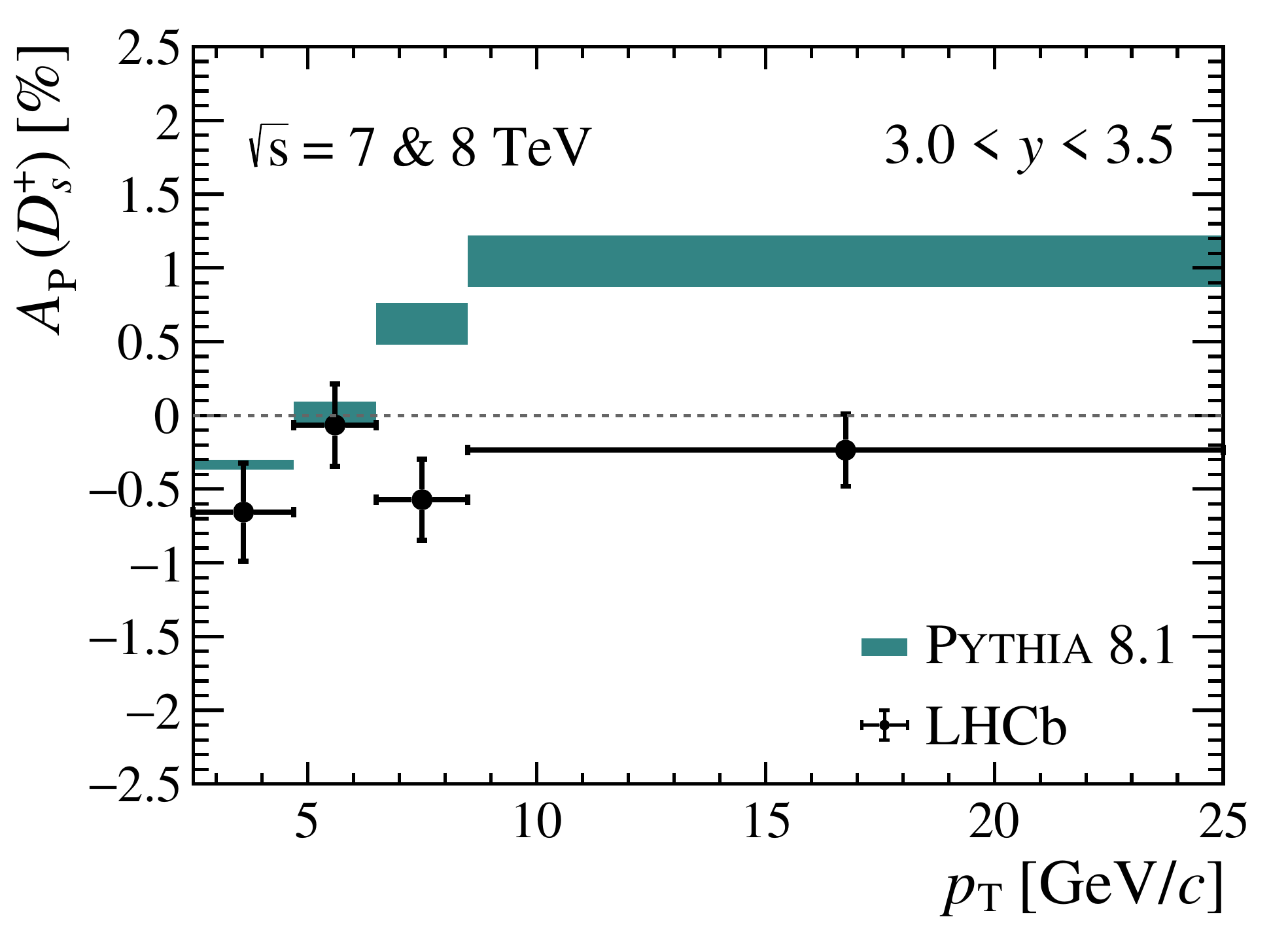}
\includegraphics[width=0.49\textwidth]{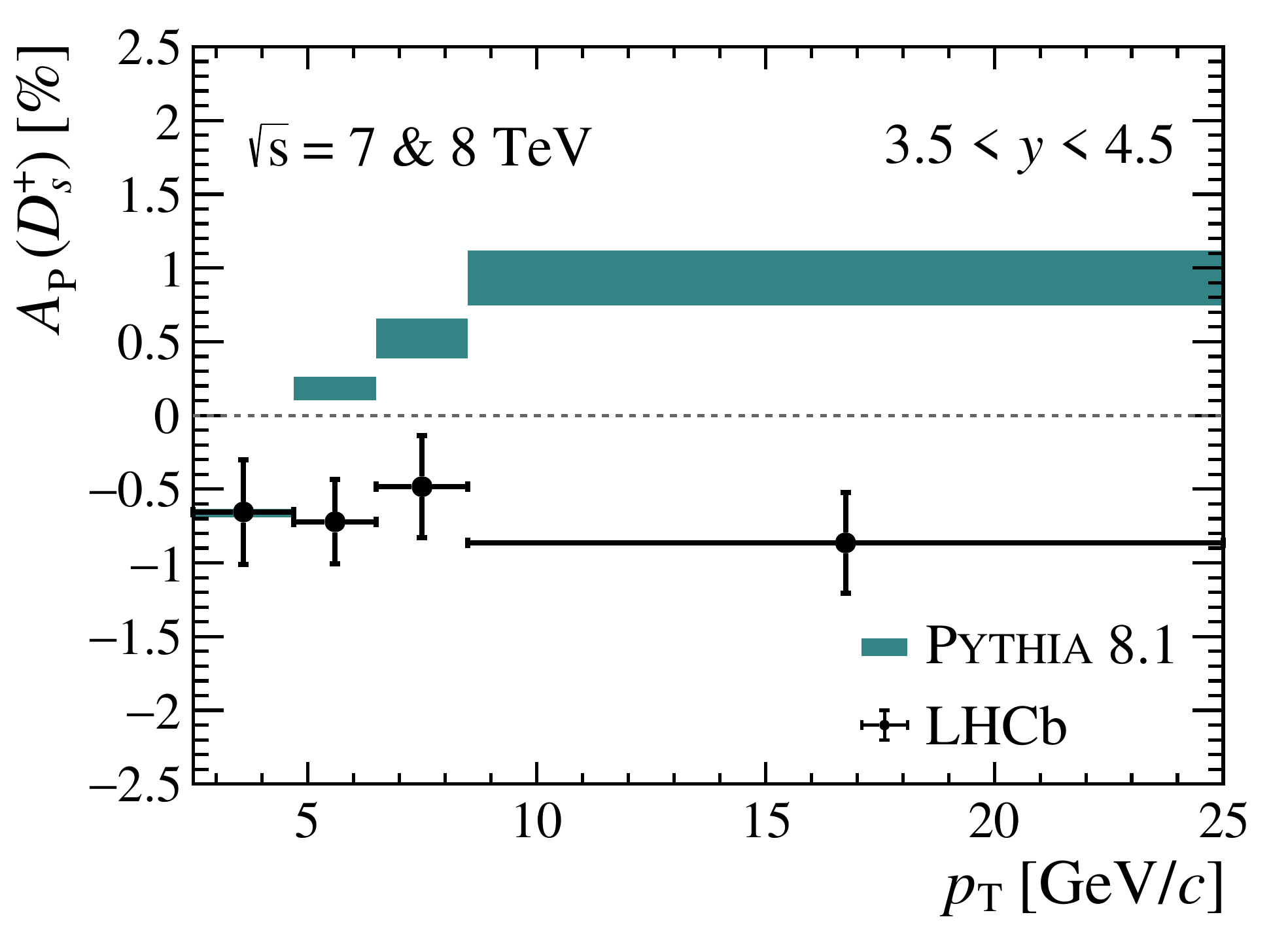}
\caption{Results of the \lhcb measurement of the \Dsp production asymmetry as 
a function of \pt for three different bins of rapidity, compared to the results 
from \pythia. Both are for the combined $\sqs=7$ and 8 TeV data sets. 
The uncertainties on the \pythia predictions are statistical only.}
\label{fig:ApDsResultsPerBin}
\end{figure}

The \pythia event generator includes models for mechanisms that cause 
production asymmetries~\cite{Norrbin:1999by, Norrbin:2000jy, Norrbin:2000zc}.
The results obtained in this paper are compared with production asymmetries
obtained from \pythia 8.1~\cite{Sjostrand:2007gs, *Sjostrand:2006za} with the 
CT09MCS set of parton density functions~\cite{Lai:2009ne}.
In this configuration, which is the default \lhcb tuning of \pythia, 
events containing a \Ds meson are extracted from generated
minimum bias interactions as described in Ref.~\cite{LHCb-PROC-2010-056}. The results of this 
comparison are shown in Fig.~\ref{fig:ApDsResultsPerBin} as a function
of \pt in the different \y bins for the combined 7 and 8\tev data sets, and separately for 7 and 8\tev
in Fig.~\ref{fig:ApDsResultsPerBin_split} in the Appendix. The \pythia simulation shows a strong
dependence on both \pt and \y, whereas the measurements presented here do not.  

\section{Summary and conclusions}
\label{sec:conclusions}

A measurement of the \Dsp production asymmetry is performed in $pp$ collisions
at centre-of-mass energies of 7 and 8\tev. The measurement is carried
out in bins of transverse momentum and rapidity, covering the range $2.5<\pt<25.0\gevc$
and $2.0<\y<4.5$, using $\Dsp\to\Kp\Km\pip$ decays, where the kaon pair is
created via the $\phi$ resonance. The production asymmetry measured in bins
of \pt and \y is shown in Fig.~\ref{fig:ApDsResultsPerBin}. 
No kinematic dependence is observed, contrary to expectations from simulations with 
the \pythia event generator. 

The results are in agreement with the previous result from the 
\lhcb collaboration~\cite{LHCb-PAPER-2012-009}, which was performed on the data recorded at
7\tev only. This updated measurement, with improvements in the detector
calibration, supersedes the previous result and provides evidence for 
a nonzero value for the production asymmetry with a significance of 3.3 standard
deviations. The results presented in this paper can be used as input to tune
the parameters of production models in different event generators.

\section*{Acknowledgements}
%
% These Acknowledgements valid from 20-Mar-2018
%
\noindent We express our gratitude to our colleagues in the CERN
accelerator departments for the excellent performance of the LHC. We
thank the technical and administrative staff at the LHCb
institutes. We acknowledge support from CERN and from the national
agencies: CAPES, CNPq, FAPERJ and FINEP (Brazil); MOST and NSFC
(China); CNRS/IN2P3 (France); BMBF, DFG and MPG (Germany); INFN
(Italy); NWO (The Netherlands); MNiSW and NCN (Poland); MEN/IFA
(Romania); MinES and FASO (Russia); MinECo (Spain); SNSF and SER
(Switzerland); NASU (Ukraine); STFC (United Kingdom); NSF (USA).  We
acknowledge the computing resources that are provided by CERN, IN2P3
(France), KIT and DESY (Germany), INFN (Italy), SURF (The
Netherlands), PIC (Spain), GridPP (United Kingdom), RRCKI and Yandex
LLC (Russia), CSCS (Switzerland), IFIN-HH (Romania), CBPF (Brazil),
PL-GRID (Poland) and OSC (USA). We are indebted to the communities
behind the multiple open-source software packages on which we depend.
Individual groups or members have received support from AvH Foundation
(Germany), EPLANET, Marie Sk\l{}odowska-Curie Actions and ERC
(European Union), ANR, Labex P2IO and OCEVU, and R\'{e}gion
Auvergne-Rh\^{o}ne-Alpes (France), Key Research Program of Frontier
Sciences of CAS, CAS PIFI, and the Thousand Talents Program (China),
RFBR, RSF and Yandex LLC (Russia), GVA, XuntaGal and GENCAT (Spain),
Herchel Smith Fund, the Royal Society, the English-Speaking Union and
the Leverhulme Trust (United Kingdom).

\clearpage

%\section*{Appendix}
\section*{Results for separate data sets}
\label{sec:App}

\begin{table}[h!]
\caption{\label{table:results2011} Values of the $\Dsp$ production asymmetry 
in percent, including, respectively, the statistical and systematic uncertainty 
for each of the \Dsp kinematic bins using the $\sqs=7\TeV$ data set.}
\begin{center}
\bgroup
\def\arraystretch{1.2}
\begin{tabular}{l @{\hskip 1.75em} *{3}{r@{ $\pm$ }l@{ $\pm$ }l } }
\toprule
& \multicolumn{9}{c}{\y} \\
\cmidrule(l{-0.5em}){2-10}
\pt [\gevc]\hskip 1.25em & \multicolumn{3}{c}{$2.0-3.0$} & \multicolumn{3}{c}{$3.0-3.5$} & \multicolumn{3}{c}{$3.5-4.5$} \\
\cmidrule[0.5pt](r{1.2em}){1-1} 
\cmidrule[0.5pt](l{-0.5em}){2-10} 
$2.5-4.7$ & $-0.74$ & $0.62$ & $0.32$ & $-1.34$ & $0.55$ & $0.13$ & $-1.15$ & $0.60$ & $0.14$ \\
$4.7-6.5$ & $-0.54$ & $0.51$ & $0.27$ & $0.16$ & $0.49$ & $0.10$ & $-0.70$ & $0.48$ & $0.13$  \\
$6.5-8.5$ & $-1.05$ & $0.40$ & $0.06$ & $-0.76$ & $0.47$ & $0.10$ & $-0.68$ & $0.56$ & $0.17$ \\
$8.5-25.0$ & $-0.14$ & $0.32$ & $0.08$ & $-0.00$ & $0.43$ & $0.10$ & $-1.18$ & $0.63$ & $0.09$ \\
\bottomrule
\end{tabular}
\egroup
\end{center}
\end{table}

\begin{table}[h!]
\caption{\label{table:results2012} Values of the $\Dsp$ production asymmetry 
in percent, including, respectively, the statistical and systematic uncertainty 
for each of the \Dsp kinematic bins using the $\sqs=8\TeV$ data set.}
\begin{center}
\bgroup
\def\arraystretch{1.2}
\begin{tabular}{l @{\hskip 1.75em} *{3}{r@{ $\pm$ }l@{ $\pm$ }l } }
\toprule
& \multicolumn{9}{c}{\y} \\
\cmidrule(l{-0.5em}){2-10}
\pt [\gevc]\hskip 1.25em & \multicolumn{3}{c}{$2.0-3.0$} & \multicolumn{3}{c}{$3.0-3.5$} & \multicolumn{3}{c}{$3.5-4.5$} \\
\cmidrule[0.5pt](r{1.2em}){1-1} 
\cmidrule[0.5pt](l{-0.5em}){2-10} 
$2.5-4.7$ & $-0.59$ & $0.40$ & $0.32$ & $-0.34$ & $0.37$ & $0.13$ & $-0.45$ & $0.39$ & $0.14$ \\
$4.7-6.5$ & $-0.73$ & $0.29$ & $0.27$ & $-0.15$ & $0.31$ & $0.10$ & $-0.73$ & $0.30$ & $0.13$ \\
$6.5-8.5$ & $-0.32$ & $0.27$ & $0.06$ & $-0.49$ & $0.31$ & $0.10$ & $-0.40$ & $0.36$ & $0.17$ \\
$8.5-25.0$ & $-0.48$ & $0.17$ & $0.08$ & $-0.32$ & $0.26$ & $0.10$ & $-0.74$ & $0.39$ & $0.09$ \\
\bottomrule
\end{tabular}
\egroup
\end{center}
\end{table}

\begin{figure}[h]
\centering
\includegraphics[width=0.49\textwidth]{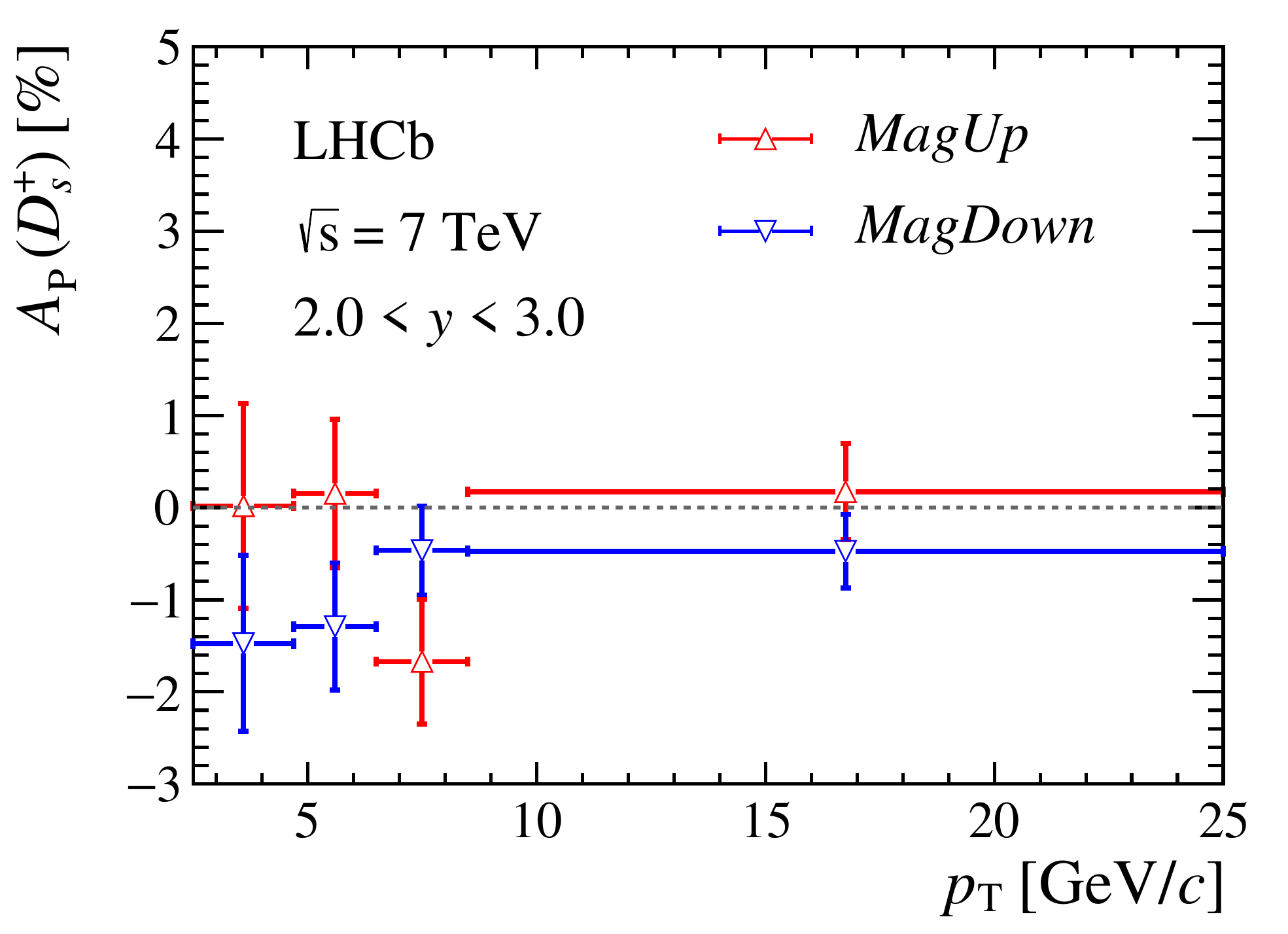}
\includegraphics[width=0.49\textwidth]{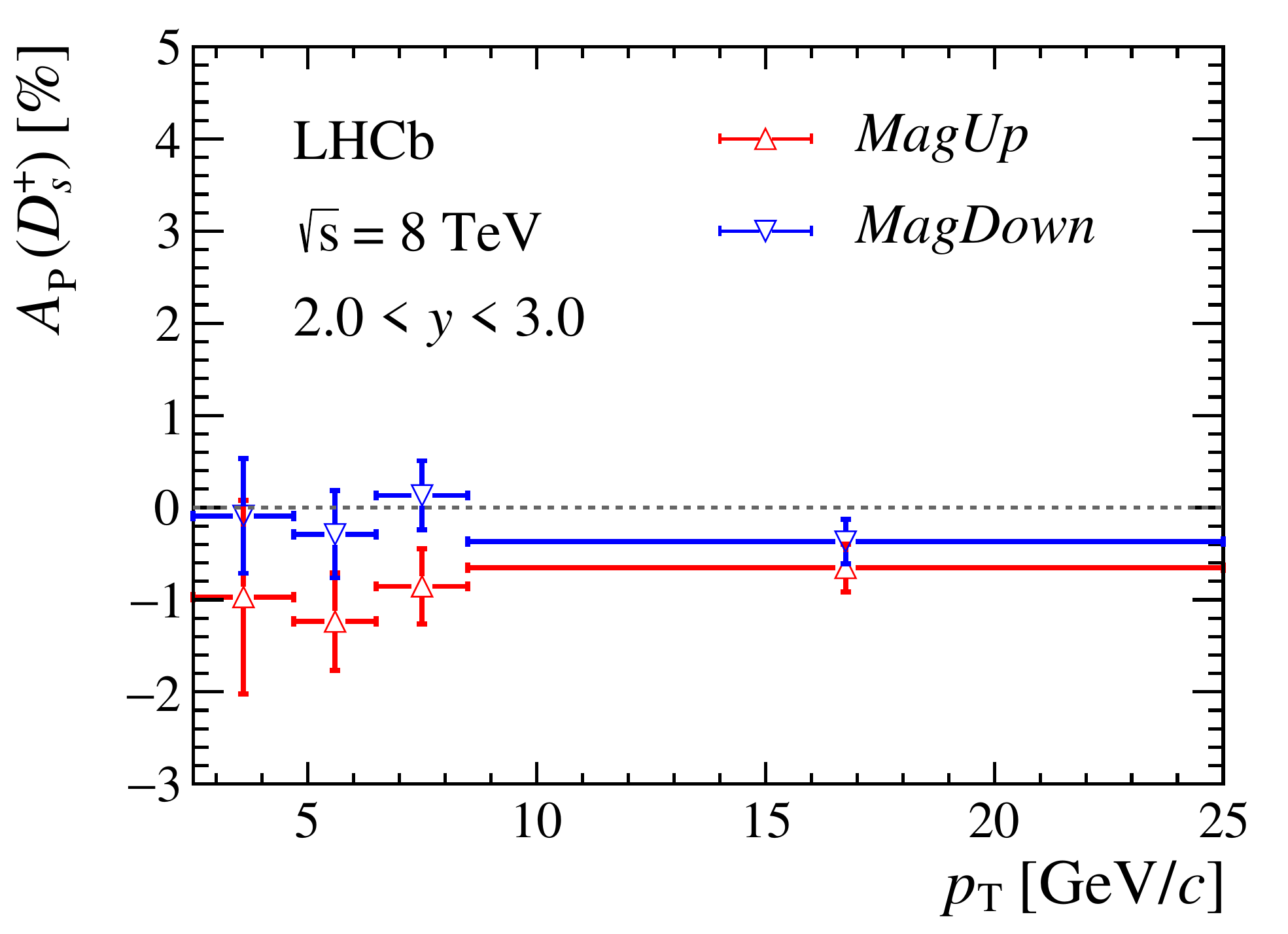}
\includegraphics[width=0.49\textwidth]{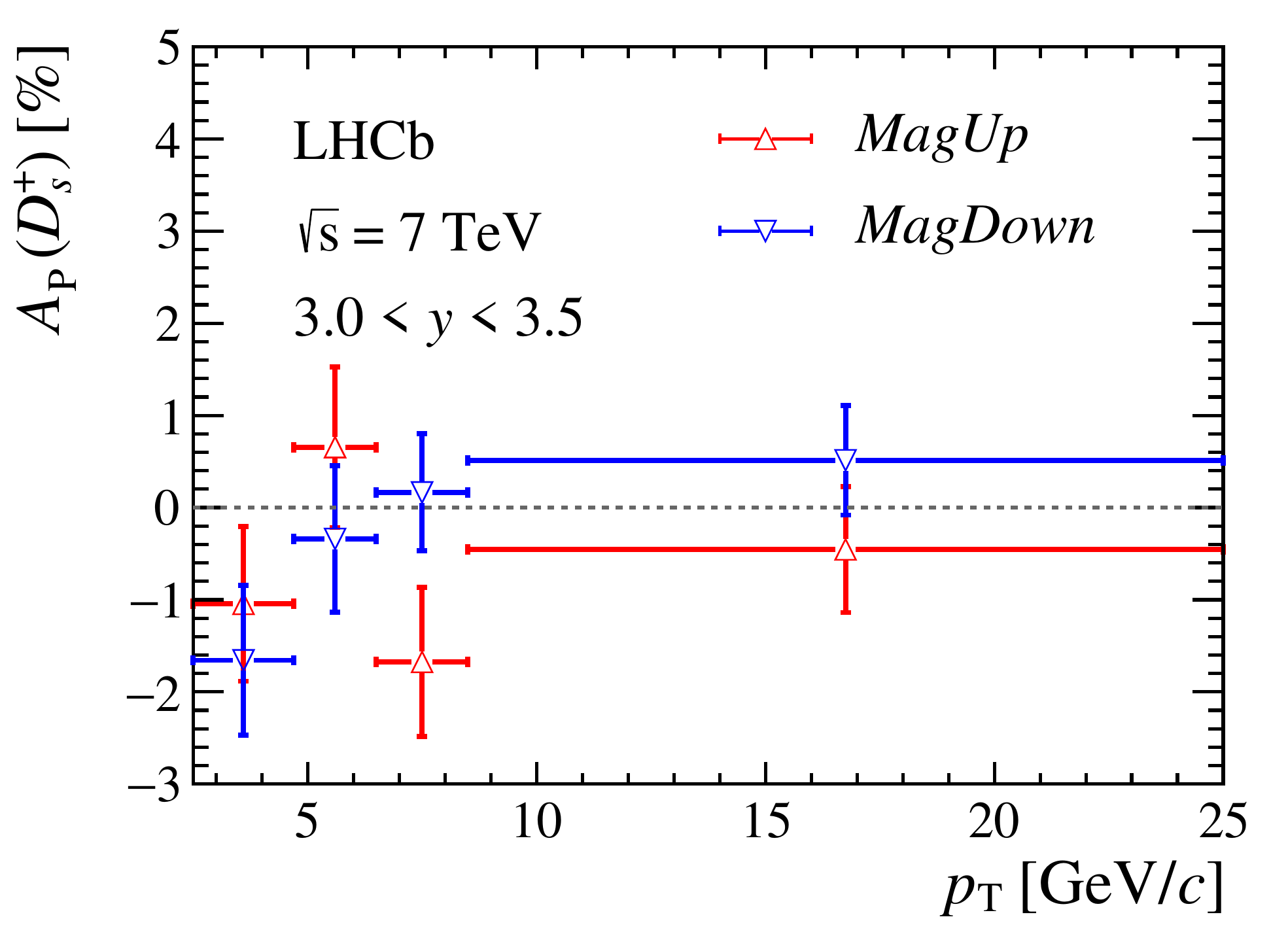}
\includegraphics[width=0.49\textwidth]{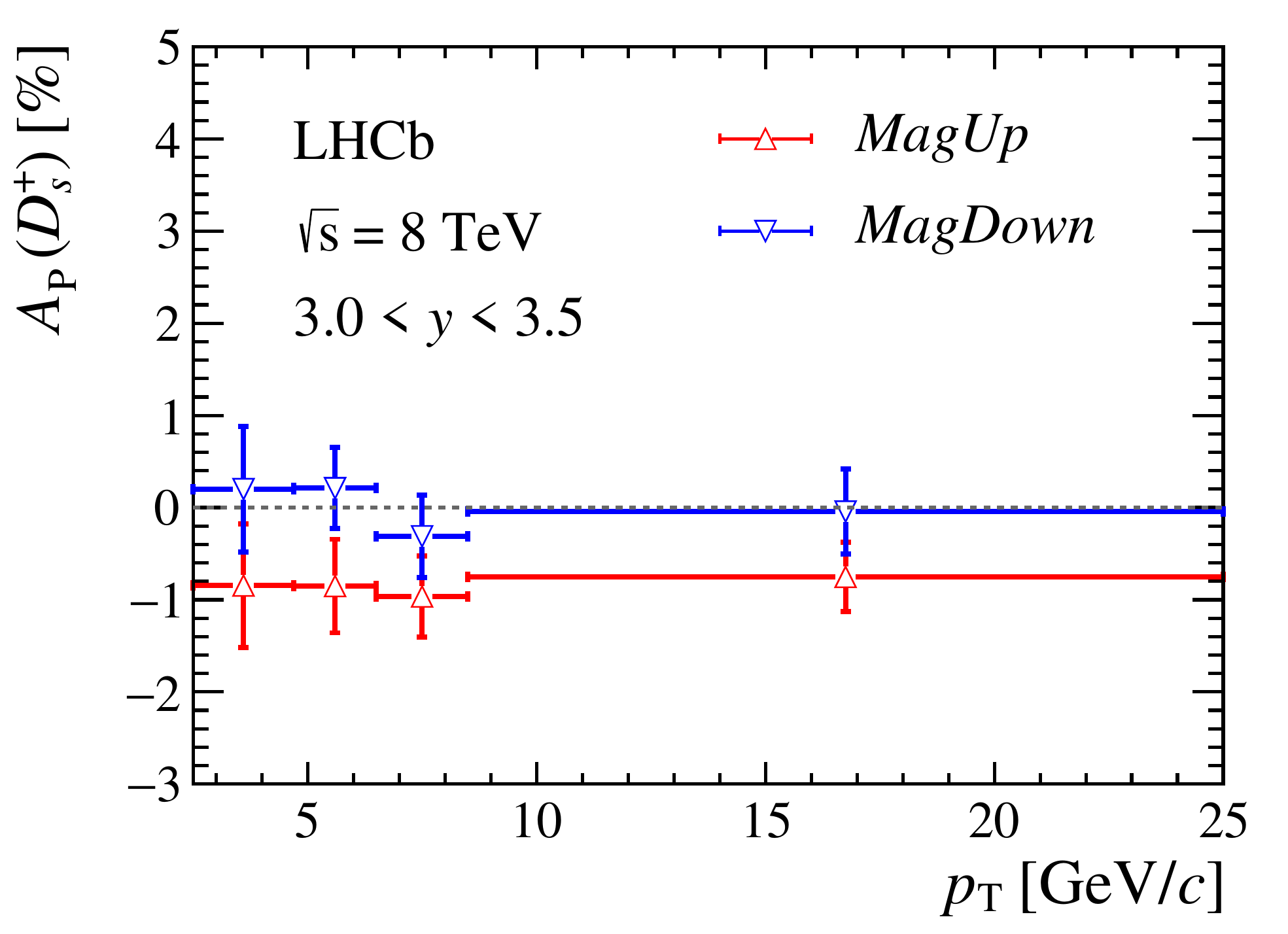}
\includegraphics[width=0.49\textwidth]{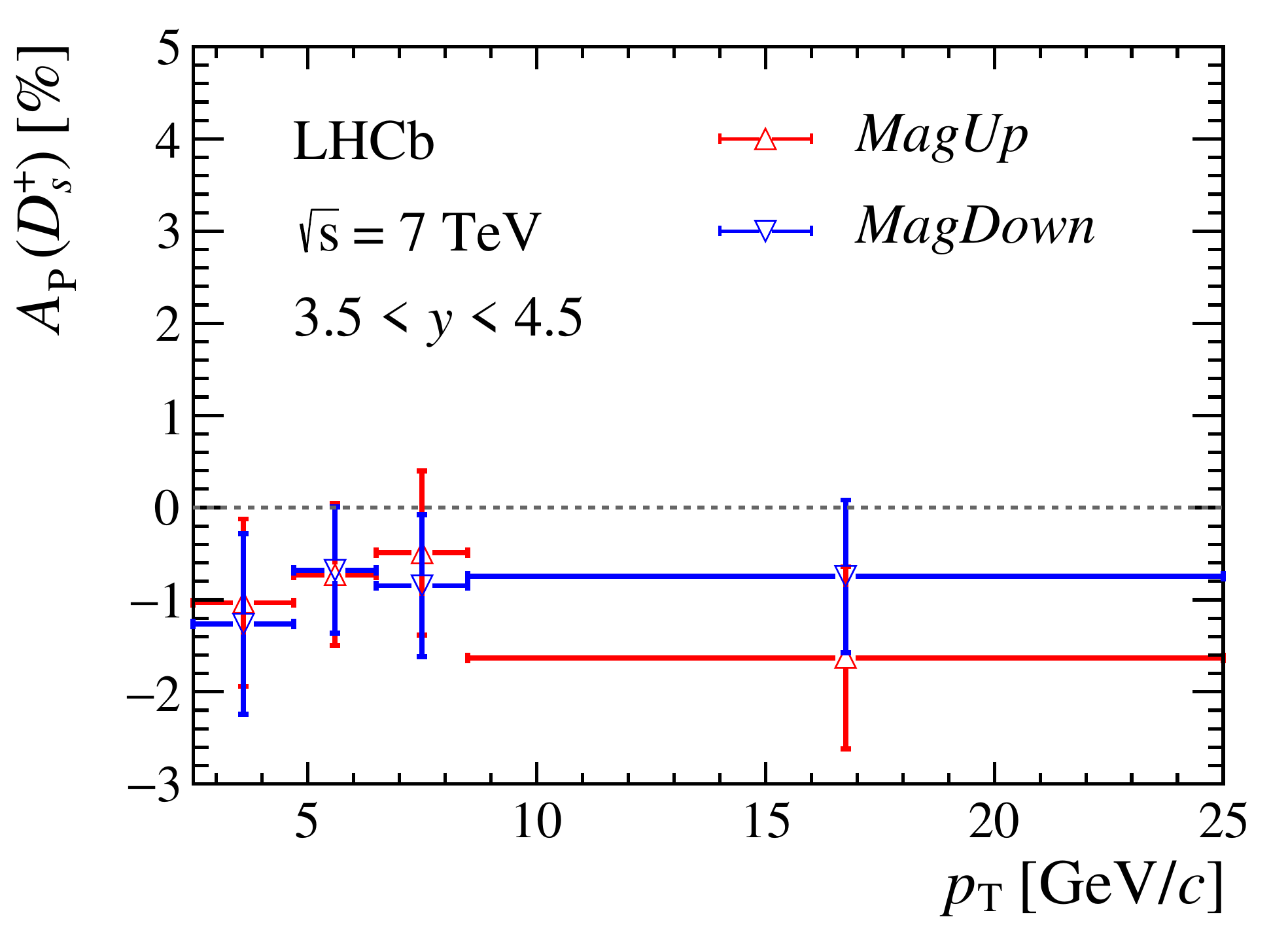}
\includegraphics[width=0.49\textwidth]{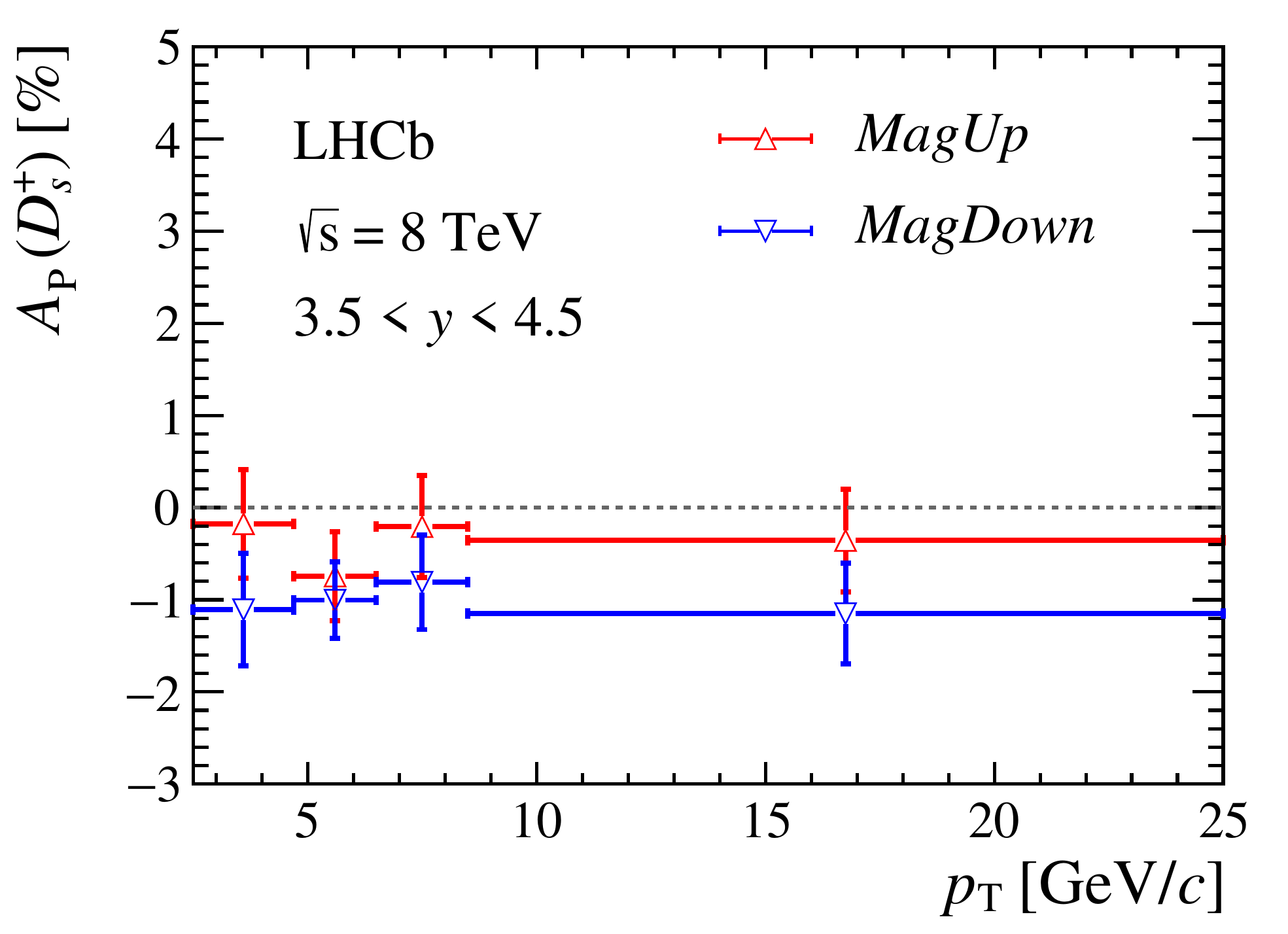}
\caption{Results of the \lhcb measurement of the \Dsp production asymmetry as 
a function of \pt for three different bins of rapidity for the (left) $\sqs=7\TeV$ and (right) $8\TeV$
data sets, split between the magnet polarities \MagUp and \MagDown.}
\label{fig:ApDsResultsPerBin_updown}
\end{figure}

\begin{figure}[h]
\centering
\includegraphics[width=0.49\textwidth]{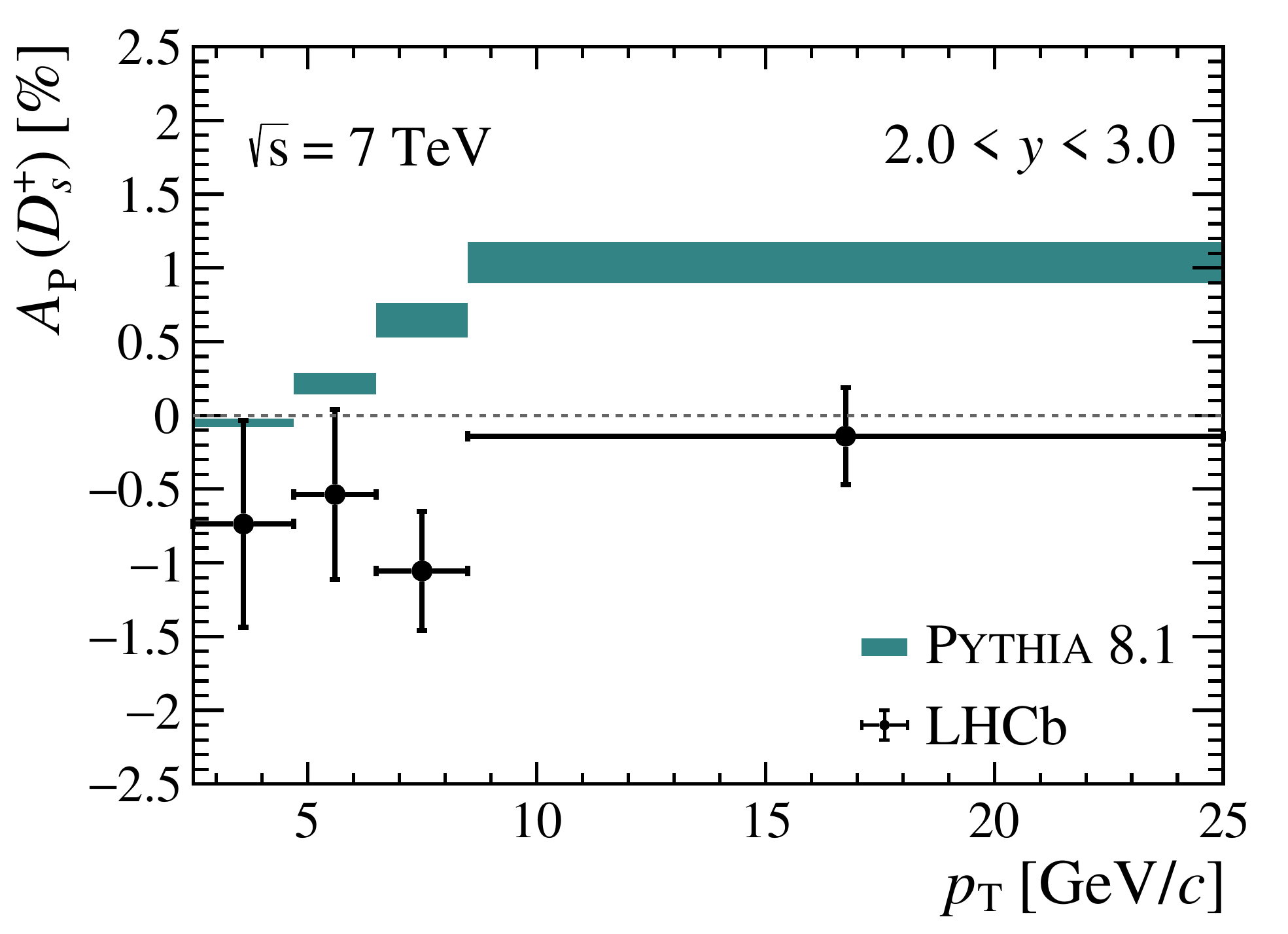}
\includegraphics[width=0.49\textwidth]{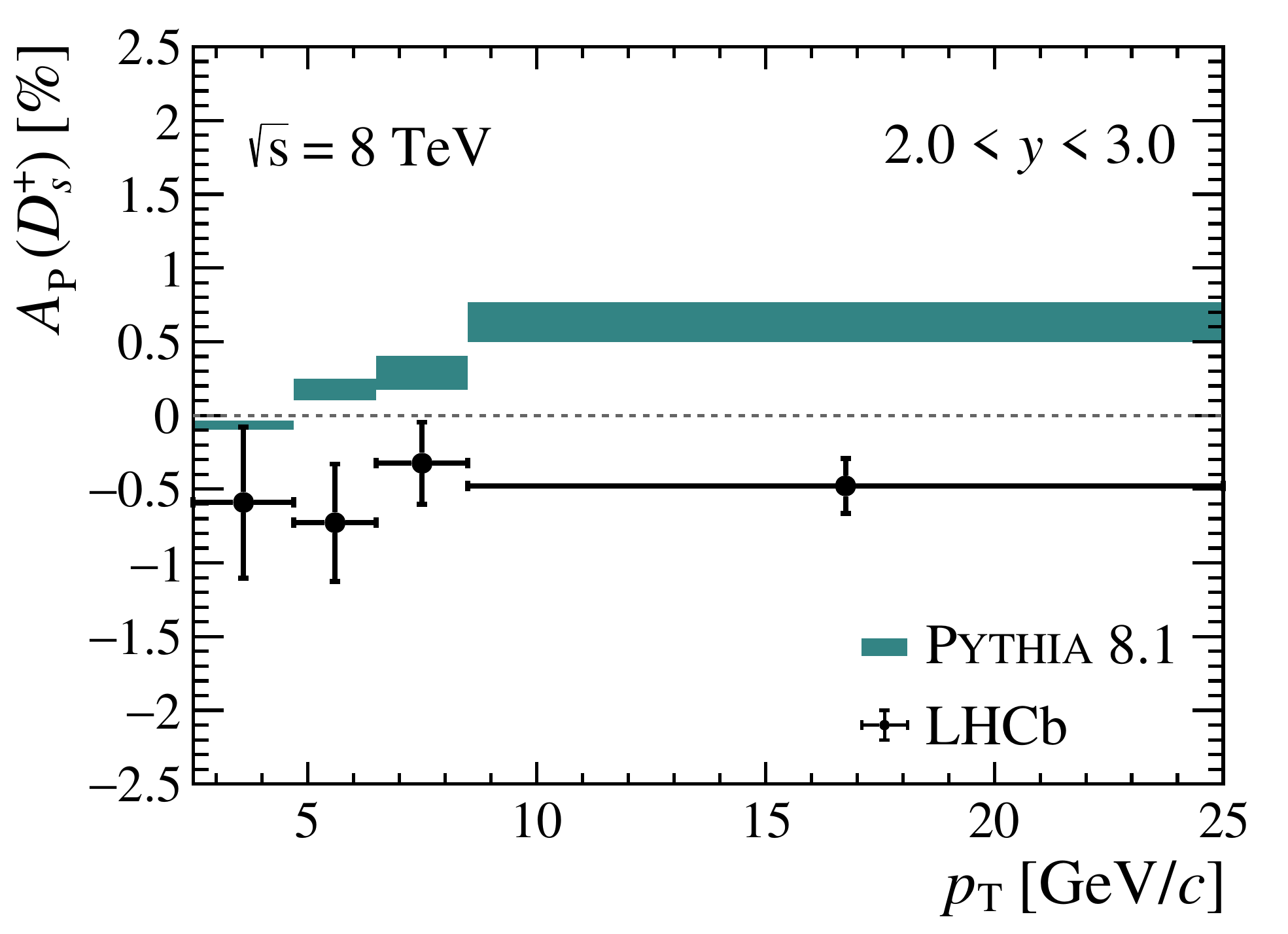}
\includegraphics[width=0.49\textwidth]{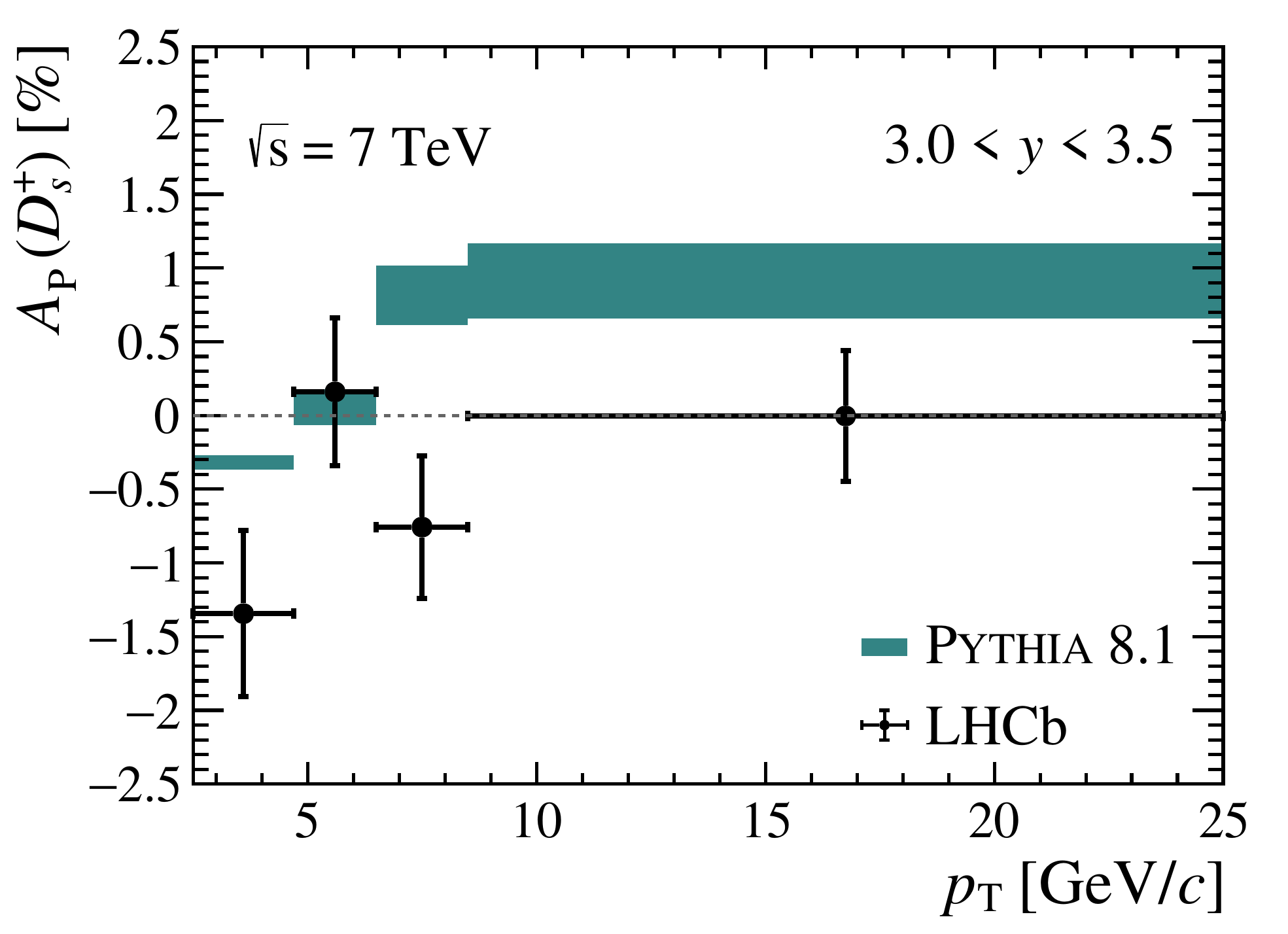}
\includegraphics[width=0.49\textwidth]{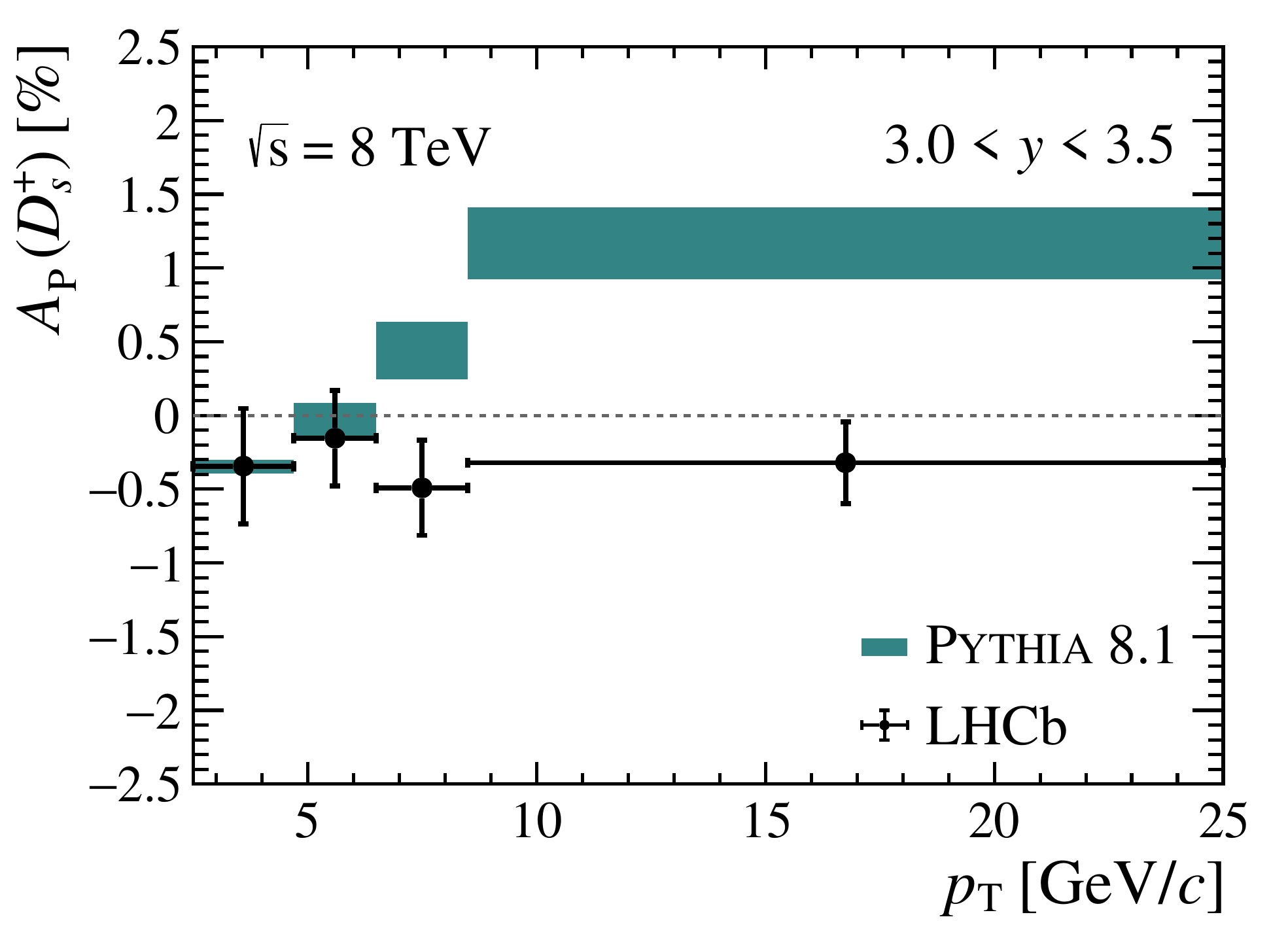}
\includegraphics[width=0.49\textwidth]{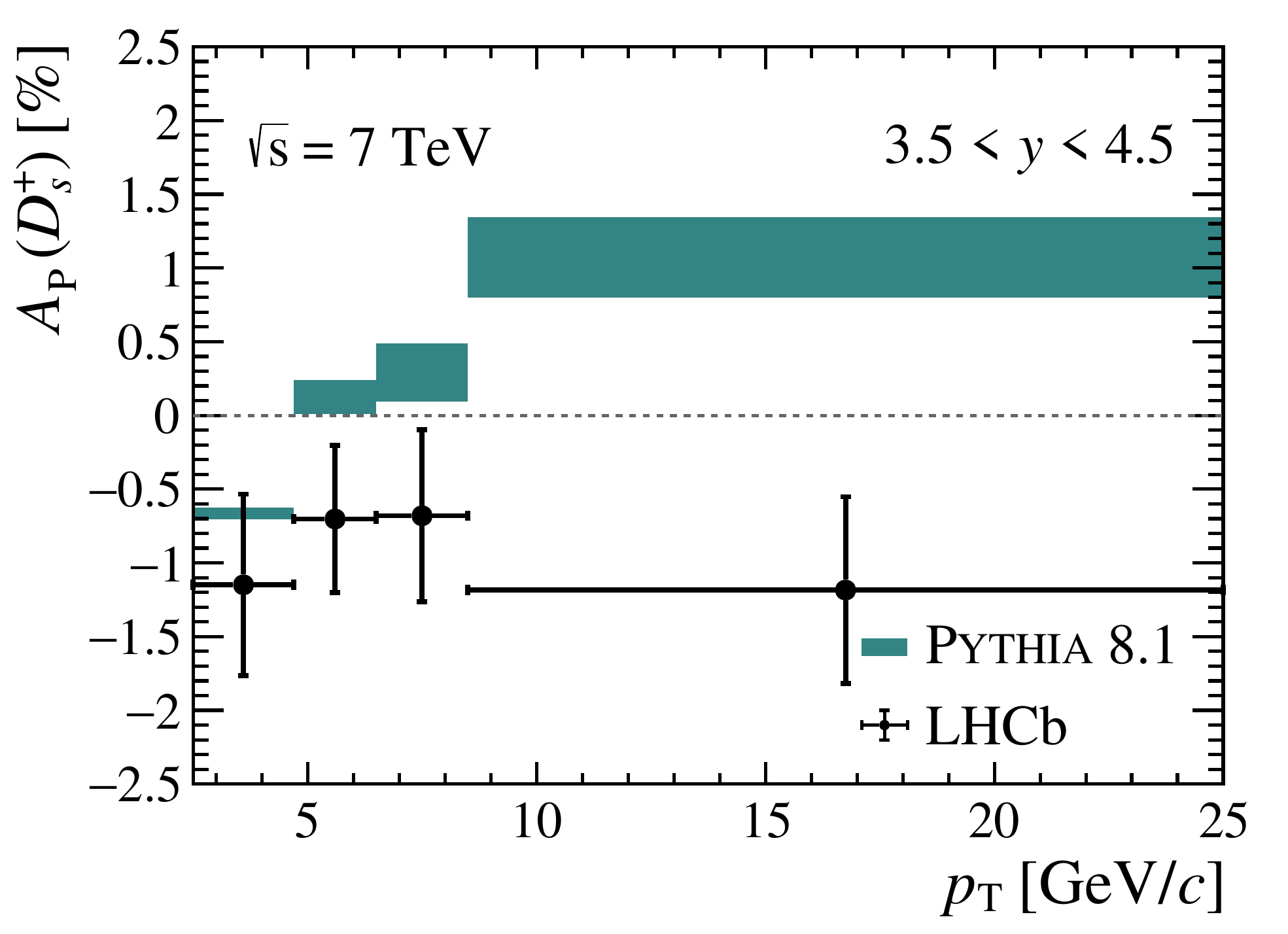}
\includegraphics[width=0.49\textwidth]{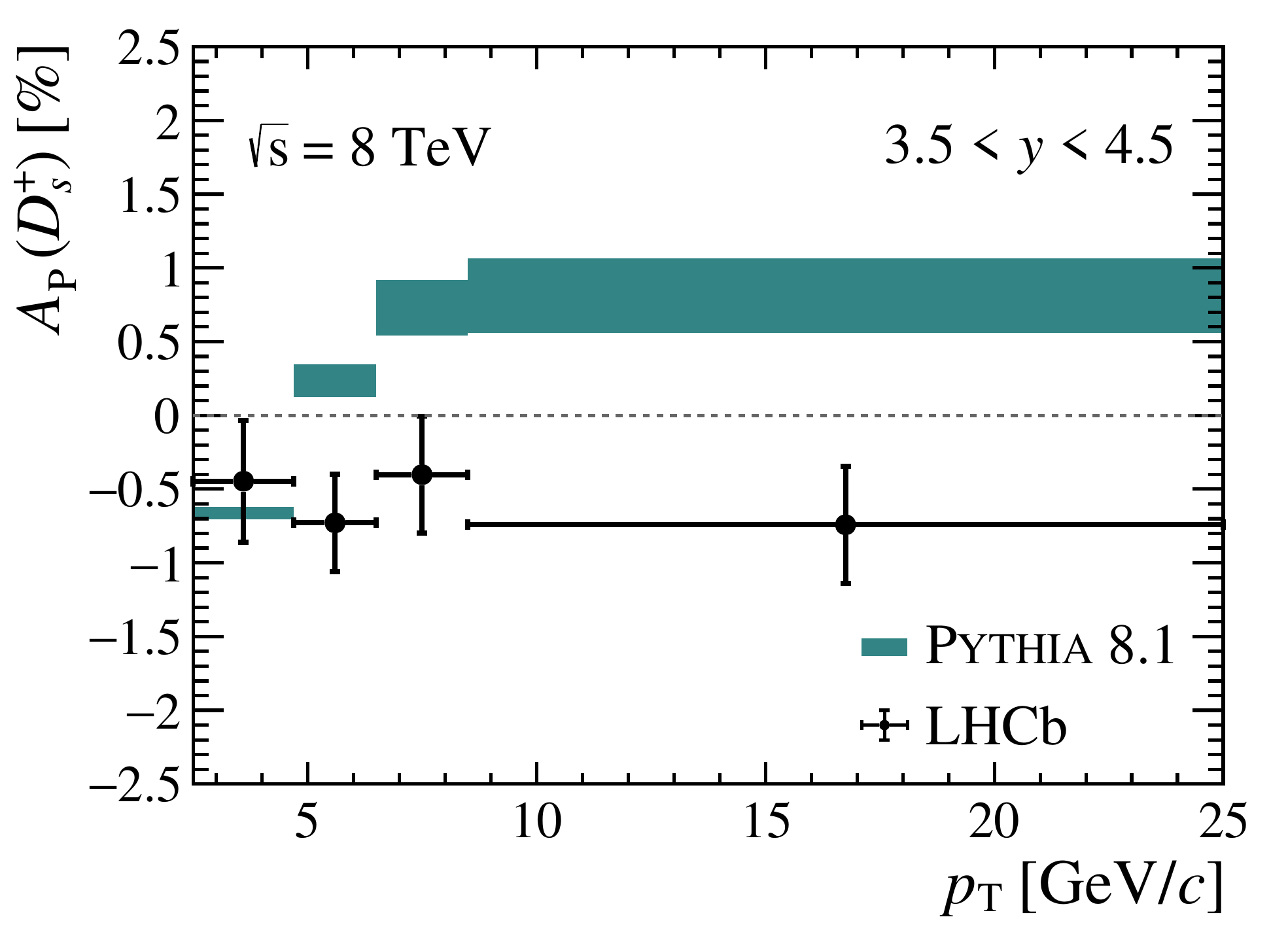}
\caption{Results of the \lhcb measurement of the \Dsp production asymmetry as 
a function of \pt for three different bins of rapidity for the (left) $\sqs=7\TeV$ and (right) $8\TeV$
data sets, compared to the results from \pythia. The uncertainties of the \pythia prediction 
are statistical only.}
\label{fig:ApDsResultsPerBin_split}
\end{figure}

\clearpage

\addcontentsline{toc}{section}{References}
\setboolean{inbibliography}{true}
\bibliographystyle{LHCb}
\bibliography{main,LHCb-PAPER,LHCb-CONF,LHCb-DP,LHCb-TDR,my-refs}

\newpage

% Author List ---------------------------- 

\newpage
\centerline{\large\bf LHCb collaboration}
\begin{flushleft}
\small
R.~Aaij$^{43}$,
B.~Adeva$^{39}$,
M.~Adinolfi$^{48}$,
Z.~Ajaltouni$^{5}$,
S.~Akar$^{59}$,
P.~Albicocco$^{18}$,
J.~Albrecht$^{10}$,
F.~Alessio$^{40}$,
M.~Alexander$^{53}$,
A.~Alfonso~Albero$^{38}$
% ...

{\footnotesize \it
% ...
$ ^{5}$Clermont Universit{\'e}, Universit{\'e} Blaise Pascal, CNRS/IN2P3, LPC, Clermont-Ferrand, France\\
% ...
$ ^{10}$Fakult{\"a}t Physik, Technische Universit{\"a}t Dortmund, Dortmund, Germany\\
% ...
$ ^{18}$INFN Laboratori Nazionali di Frascati, Frascati, Italy\\
% ...
$ ^{38}$ICCUB, Universitat de Barcelona, Barcelona, Spain\\
$ ^{39}$Instituto Galego de F{\'\i}sica de Altas Enerx{\'\i}as (IGFAE), Universidade de Santiago de Compostela, Santiago de Compostela, Spain\\
$ ^{40}$European Organization for Nuclear Research (CERN), Geneva, Switzerland\\
% ...
$ ^{43}$Nikhef National Institute for Subatomic Physics, Amsterdam, The Netherlands\\
% ...
$ ^{48}$H.H. Wills Physics Laboratory, University of Bristol, Bristol, United Kingdom\\
% ...
$ ^{53}$School of Physics and Astronomy, University of Glasgow, Glasgow, United Kingdom\\
% ...
$ ^{59}$University of Cincinnati, Cincinnati, OH, United States\\
% ...
}
\end{flushleft}

\end{document}